\documentclass[reprint,amsmath,amssymb,aps]{revtex4-2}
\usepackage[colorlinks,citecolor=blue,linkcolor=blue,anchorcolor=blue,filecolor=blue,
urlcolor=blue]{hyperref}
\usepackage{graphicx}
\usepackage{ulem}
\usepackage{dcolumn}
\usepackage{bm}
\usepackage{todonotes}
\usepackage{bigints}
\usepackage{amsmath}

\newcommand{\kin}{\mathrm{kin}}
\newcommand{\ma}{\mathrm{max}}
\newcommand{\mi}{\mathrm{min}}
\newcommand{\ec}{\mathrm{ec}}
\newcommand{\masy}{\mathrm{ma}}
\newcommand{\inp}{\mathrm{inp}}
\newcommand{\sky}{\mathrm{sky}}
\newcommand{\gog}{\mathrm{gog}}
\newcommand{\sym}{\mathrm{sym}}
\newcommand{\sm}{\mathrm{SNM}}
\newcommand{\pot}{\mathrm{pot}}
\newcommand{\corr}{\mathrm{corr}}

\begin{document}

\title{Asymmetric warm nuclear matter described by Gogny and Skyrme-version models}
\author{Odilon Louren\c{c}o$^1$, Mariana Dutra$^1$, Mario Centelles$^{2,3}$, and Xavier Vi\~nas$^{2,3,4}$}
\affiliation{
$^1$Departamento de F\'isica e Laborat\'orio de Computa\c c\~ao Cient\'ifica Avan\c cada e Modelamento (Lab-CCAM), Instituto Tecnol\'ogico de Aeron\'autica, DCTA, 12228-900, S\~ao Jos\'e dos Campos, SP, Brazil
\\
$^2$Departament de F\'isica Qu\`antica i Astrof\'isica (FQA), Universitat de Barcelona (UB), Mart\'i i Franqu\`es 1, E-08028 Barcelona, Spain
\\
$^3$Institut de Ci\`encies del Cosmos (ICCUB), Universitat de Barcelona (UB), Mart\'i i Franqu\`es 1, E-08028 Barcelona, Spain
\\
$^4$Institut Menorqu\'i d'Estudis, Cam\'i des Castell 28, E-07702 Ma\'o, Spain
}

\date{\today}

\begin{abstract}
In this work, we perform a detailed study of the thermodynamical properties of asymmetric nuclear matter at finite temperatures by means of the Gogny force, with a particular focus on its D1 family. We emphasize the investigation of the liquid-gas phase transition with the respective analysis of phase-coexistence boundaries (binodal sections) and instability regions (spinodal contours). Furthermore, the phenomenon of isospin distillation, intrinsically related to the unstable part of the system, is also described. In order to estimate the impact of the finite range of the nuclear interaction, for each Gogny parametrization we provide a respective zero-range Skyrme version, for which the free parameters of the model are obtained with the aim of reproducing
at zero temperature the same saturation density, binding energy, incompressibility, isoscalar effective mass, isovector effective mass, symmetry energy, and symmetry energy slope. As a main result, we verify systematic deviations between the Gogny models and their Skyrme-version models, particularly at higher temperatures, where the Skyrme-version parametrizations exhibit reduced binodal and spinodal regions.
\end{abstract}

\maketitle

\section{Introduction}

The study of nuclear matter properties as a function of the density, temperature, and isospin asymmetry is a fundamental problem in nuclear physics. The main source of experimental information about this system is provided by heavy ion collisions and highly excited nuclei created by radioactive beams~\cite{muller95}. Some theoretical analyses are addressed to study the equilibrium thermodynamics paying special attention to the liquid-gas phase transition at low densities and moderate temperatures investigating how the nuclear system evolves through the phase-separation (binodal) and instability (spinodal) boundaries~\cite{muller95}.

From a thermodynamical point of view, a macroscopic heated liquid is surrounded by the emitted vapor, and the stability of this system is reached if the inside liquid phase and the outside vapor phase are in thermodynamical equilibrium. In a microscopic compound nucleus description, one shall also consider the hot nucleus in equilibrium with the gas of evaporated nucleons. As temperature increases, the vapor contribution also increases up to reaching a limiting temperature beyond which the compound nucleus ceases to be a stable bound system~\cite{de90}. In the nuclear matter approach, this translates into the existence of a critical temperature beyond which the van der Waals~(vdW) pattern, responsible for the high and the low-density phases' coexistence, disappears. Such a vdW pattern can be observed, for instance, in the pressure ($P$) versus density ($\rho$) curve, in which at very low densities one has $P\sim \rho$ (ideal gas behavior). As $\rho$ increases, the attractive interaction induces a decrease of $P$ until the repulsive interaction becomes dominant at higher densities, again causing an increase in pressure. The critical temperatures computed with different mean-field models lie in the range between $13$ and $24$~MeV (see~\cite{wang05} and references therein).  In a recent paper~\cite{dutra2021}, some of us have collected a wide catalog of critical points of symmetric nuclear matter obtained with different zero-range Skyrme and finite-range Gogny, M3Y, Momentum Dependent Interaction (MDI) and Simple Effective Interaction (SEI).

The thermodynamical properties of hot asymmetric nuclear matter in the nonrelativistic domain have been discussed from long ago using zero-range Skyrme forces \cite{lattimer78,lamb81,rayet82,lattimer85,brack85,pethick95,sil04} or a simple energy density functional \cite{barranco80}. There are also studies of hot asymmetric nuclear matter using finite-range forces available in the literature. Let us mention in this respect the calculations of Refs.~\cite{heyer88,kurchareck89,hong90,zhang96} using Gogny forces, ~\cite{csernai92,mishra93,li97,xu07,moustakidis2008} with MDI, \cite{behera02,behera09,behera11}
with SEI and \cite{de90,moshfegh11} with the Seyler-Blanchard force.

Our aim in this work is twofold. On the one hand, we want to analyze in detail the thermal properties in asymmetric nuclear matter predicted by the Gogny forces of the D1 family \cite{Decharge1980,Berger1991,PLB668Chappert2008,PRL102Goriely2009}.
In particular, we will study the phase coexistence (binodal) and the instability regions (spinodals) as a function of the temperature because this information, to our knowledge, has not been discussed in detail in earlier literature using Gogny forces. On the other hand, we want to estimate the impact of the finite range of the effective interaction on the thermal properties of asymmetric nuclear matter. To this end, we derive a zero-range Skyrme interaction where the free constants of this model are fixed in order to reproduce several nuclear empirical parameters predicted by the Gogny force at zero temperature. The thermal predictions in asymmetric nuclear matter from the Skyrme-version model compared to the ones predicted by the genuine Gogny force allow us to verify the influence of the finite range in hot asymmetric nuclear matter.

We organize our work as follows. In Sec.~\ref{sec:hotmod} we present the structure of the main equations of state, at finite temperature, for the Gogny model and its respective Skyrme-version. We dedicate Sec.~\ref{sec:results} for the presentation of our analysis concerning the two-component asymmetric matter at finite temperature, with results related to the liquid-gas phase transition exhibited at a moderate range of temperature. Effects coming from the finite range interaction on the binodal and spinodal regions of the phase diagrams are also described at this point. The stability of the asymmetric matter against separation into two phases and the isospin distillation effect is also analyzed.
Finally, a summary and our concluding remarks are shown in Sec.~\ref{sec:summ}.

\section{Gogny and Skyrme-like models at finite temperature}
\label{sec:hotmod}

To describe self-consistently asymmetric nuclear matter one starts minimizing the thermodynamical potential \cite{heyer88,de90}
\begin{equation}
\Omega(\mu_n,\mu_p,T) = E - TS - \mu_n N - \mu_p Z
\label{eq1}
\end{equation}
with respect to the occupation numbers of each type of nucleon $n_q$ ($q=n,p$). In Eq.~(\ref{eq1}), $E$ and $S$ are the total energy and entropy and $\mu_n$ and $N$ ($\mu_p$ and $Z$) are the neutron (proton) chemical potential and particle number, respectively.

In the Landau quasi-particle approximation, the entropy density for each type of nucleon can be written as
\begin{eqnarray}
{\cal{S}}_q &=& - \sum_k \Big\{ n_q(k,T)\ln[n_q(k,T)]
\nonumber \\
&& \mbox{} + [1 - n_q(k,T)] \ln[1 - n_q(k,T)] \Big\} .
\label{eq2}
\end{eqnarray}
Using (\ref{eq2}) in (\ref{eq1}), the occupation numbers $n_q(k,T)$, which are the variational solution of Eq.~(\ref{eq1}), are given by the Fermi-Dirac distribution
\begin{equation}
n_q(k,T) = F_{Dq}(k,T) = \left\{1 + e^{[\varepsilon_q(k) - \mu_q]/T}\right\}^{-1},
\label{eq3}
\end{equation}
where $\varepsilon_q(k)$ are the single-particle energies written as
\begin{equation}
\varepsilon_q(k) = \frac{\hbar ^2k^2}{2M} + u_q (k,\rho_n,\rho_p),
\label{eq4}
\end{equation}
being $u_q (k,\rho_n,\rho_p)$ the single-particle potential that depends on the mean-field model, and $M$ the bare nucleon mass.
The $u_q (k,\rho_n,\rho_p)$ functions can be extracted from the Gogny and Skyrme models presented below. The nucleon densities $\rho_q$ ($q=n,p$) are given by:
\begin{equation}
\rho_q = 2 \int\!\!\frac{d\bf{k}}{(2\pi)^3} \, F_{Dq}(k,T) ,
\label{eq:dens}
\end{equation}
where the summation over momenta has been transformed in the usual fashion into a continuous integration, and the multiplying factor $2$ accounts for spin degeneracy.

\subsection{The Gogny model}
The finite-range effective Gogny interaction of the D1 family has the following structure
\cite{Decharge1980,arnau1,claudia2017}
\begin{eqnarray}
v(\bf{r_1},\bf{r_2})&=& \sum_{i=1}^{2} ( W_i + B_i P_{\sigma} -  H_i P_{\tau} - M_i  P_{\sigma} P_{\tau}) e^{-r^2/\mu_i^2}
\nonumber \\
&+& t_3 (1 + P_{\sigma}) \rho^{\alpha}(\bf{R}) \delta(\bf{r}),
\label{eq5}
\end{eqnarray}
where $\bf{r}=\bf{r_1}-\bf{r_2}$ and $\bf{R}=(\bf{r_1}+\bf{r_2})/2$ are the relative and center of mass coordinates, $W_i$, $B_i$, $H_i$, and $M_i$ are the strengths of the possible combinations of spin ($P_{\sigma}$) and isospin ($P_{\tau}$) exchange operators, respectively, $\mu_i$ are the ranges of the two Gaussian form factors, and $\alpha$ gives the density dependence of the two-body zero-range part that simulates the three-body interaction. Concretely, for the calculations of this work we will employ the  Gogny parametrizations D1S \cite{Berger1991}, D1N \cite{PLB668Chappert2008}, and D1M \cite{PRL102Goriely2009}. In Eq.~(\ref{eq5}) we have neglected the spin-orbit and tensor terms of the interaction owing to the fact that they do not contribute to the energy density in infinite nuclear matter.

In the asymmetric nuclear matter case, the single-particle energy for each type of nucleon is given by
\begin{align}
&\varepsilon_q(k) = \frac{\hbar^2 k^2}{2M} + V_{\rm zero\mbox{-}range},q(\rho_q,\rho_{\bar{q}})
+ V_{\rm dir,q}(\rho_q,\rho_{\bar{q}})
\nonumber\\
&+ V_{\rm exch,q}(k, \mu_n, \mu_p)
\nonumber\\
&= \frac{\hbar^2 k^2}{2M} + \frac{t_3}{4}
\left[\rho^{-2/3}(\rho^2-\rho_q^2 -\rho_{\bar{q}}^2) + 6 \rho^{1/3} \rho_{\bar{q}})\right]
\nonumber\\
&+ \sum_{i=1}^2 {\tilde g}(0,0,\mu_i)\Bigg[\left(W_i + \frac{B_i}{2} - M_i -  \frac{H_i}{2}\right)\rho_q
\nonumber\\
&+ \left(W_i + \frac{B_i}{2}\right)\rho_{\bar{q}}\Bigg] + \gamma \sum_{i=1}^2\Bigg[ \left(M_i + \frac{H_i}{2} - B_i -  \frac{W_i}{2}\right)
\nonumber\\
&\times \int\!\!\frac{d\bf{k'}}{(2\pi)^3}F_{Dq}(k',T) \, {\tilde g}(k,k',\mu_i)
 + \left(M_i + \frac{H_i}{2}\right)
\nonumber\\
&\times \int\!\!\frac{d\bf{k'}}{(2\pi)^3}F_{D\bar{q}}(k',T) \, {\tilde g}(k,k',\mu_i) \Bigg],
\label{eq6}
\end{align}
where the total density is $\rho=\rho_n+\rho_p$, $\gamma$ is the degeneracy factor (2 in the case of asymmetric nuclear matter),
and, for the specified isospin-index $q$, ${\bar{q}}$ describes the other isospin-index.
The $F_{Dq}(k,T)$ function is given by Eq.~(\ref{eq3}) and ${\tilde g}(k,k',\mu_i)$ is the angular-averaged Fourier transform of the finite-range form-factor with range $\mu_i$ (see Appendix of Ref.~\cite{dutra2021}).
In the case of Gogny forces, ${\tilde g}(k,k',\mu_i)$ reads as follows:
\begin{align}
{\tilde g}(k,k',\mu_i)= \frac{2 \pi^{3/2}\mu_i}{k k'}e^{-\mu_i^2(k^2 + {k'}^2)/4} \sinh\left( \frac{\mu_i^2 k k'}{2}\right).
\label{eq7}
\end{align}

In contrast to the zero-range forces (see below), in order to determine the chemical potentials $\mu_q^{\gog}$ ($q=n,p$) for given neutron and proton densities, one has to solve the set of four coupled equations formed by the integrals of the occupation numbers for each type of particle together with the single-particle energies (\ref{eq6}) for neutrons and protons. Once the occupation numbers are determined, the energy density and entropy density can be computed, respectively, as
\begin{widetext}
\begin{eqnarray}
{\cal H}^{\gog} &=& \frac{\hbar^2}{2M} \int\!\!\frac{d\bf{k}}{(2\pi)^3} [F_{Dn}(k,T) + F_{Dp}(k,T)]k^2
+ \frac{3}{4}t_3 \rho^{1/3} \left(\rho^2 - \rho_n^2 - \rho_p^2\right)
\nonumber \\
&+&
\frac{1}{2} \sum_{q=n,p}\sum_{i=1}^2 {\tilde g}(0,0,\mu_i)\left[\left(W_i + \frac{B_i}{2} - M_i -  \frac{H_i}{2}\right)\rho_q^2
+ \left(W_i + \frac{B_i}{2}\right)\rho_q\rho_{\bar{q}}\right]
\nonumber \\
&+&
\frac{\gamma^2}{2} \sum_{q=n,p}\sum_{i=1}^2 \left[\left(M_i + \frac{H_i}{2} - B_i -  \frac{W_i}{2}\right)
\int\!\!\frac{d\bf{k}}{(2\pi)^3}F_{Dq}(k,T) \int\!\!\frac{d\bf{k'}}{(2\pi)^3}F_{Dq}(k',T) \, {\tilde g}(k,k',\mu_i) \right.
\nonumber \\
&+& \left.\left(M_i + \frac{H_i}{2}\right) \int\!\!\frac{d\bf{k}}{(2\pi)^3}F_{Dq}(k,T) \int\!\!\frac{d\bf{k'}}{(2\pi)^3}F_{D\bar{q}}(k',T) \, {\tilde g}(k,k',\mu_i) \right],
\label{eq8}
\end{eqnarray}
\end{widetext}
and
\begin{align}
&{\cal S}^{\gog} =
\nonumber\\
&-\frac{2}{T}\sum_{q=n,p} \int\!\!\frac{d\bf{k}}{(2\pi)^3}F_{Dq}(k,T) \left[\varepsilon_q(k) - \mu_q + \frac{k}{3} \frac{d\varepsilon_q(k)}{dk}\right].
\label{eq9}
\end{align}
In particular, analytical expressions for $\mu_p^{\gog}$ and $\mu_n^{\gog}$ at $T=0$ can be found in Ref.~\cite{claudia2017}.

The pressure at a given temperature $T$ is given by the standard thermodynamical relation
\begin{equation}
P^{\gog}(\rho_n,\rho_p,T)=\mu_n^{\gog} \rho_n + \mu_p^{\gog} \rho_p - {\cal H}^{\gog} + T {\cal S}^{\gog},
\label{eq11}
\end{equation}
that can also be used to extract the Helmholtz free energy density, defined as ${\cal F}^{\gog}= {\cal H}^{\gog} - T{\cal S}^{\gog}$.

Another interesting property is the nucleon effective mass, which characterizes the momentum dependence of the nucleon single-particle potential. It is obtained as~\cite{mahaux89,shang2020,vinas21}
\begin{equation}
\frac{\hbar^2 k}{M_q^{\gog *}} = \frac{\partial \varepsilon_q(k)}{\partial k} = \frac{\hbar^2 k}{M} + \frac{\partial V_{\rm exch,q}(k, \mu_n, \mu_p)}{\partial k},
\end{equation}
which for the Gogny model at finite temperature leads to
\begin{align}
f_q^{\gog}&=\frac{M}{M_q^{\gog *}} = 1 + \frac{2M}{\hbar^2} \frac{1}{2k} \frac{\partial V_{\rm exch,q}(k, \mu_n,\mu_p)}{\partial k}
\nonumber\\
&= 1 + \frac{2M}{\hbar^2} \frac{\gamma}{2k}\sum_{i=1}^{2} \Bigg[ 
\left( M_i + \frac{H_i}{2} - B_i - \frac{W_i}{2} \right)\times
\nonumber\\
&\times\int \frac{d\bf{k'}}{(2\pi)^3}F_{Dq}(k',T) \frac{\partial \tilde{g}(k, k', \mu_i)}{\partial k} 
\nonumber\\
&+\left( M_i + \frac{H_i}{2} \right) 
\int \frac{d\bf{k'}}{(2\pi)^3}F_{D\bar{q}}(k',T) \frac{\partial \tilde{g}(k, k', \mu_i)}{\partial k}\Bigg],
\label{eq:mgog}
\end{align}
where
\begin{align}
&\frac{\partial \tilde{g}(k, k', \mu_i)}{\partial k} = \frac{\pi^{3/2} \mu_i}{k^2 k'} e^{-\mu_i^2 (k^2 + k'^2)/4}\times
\nonumber\\
&\times\Bigg[ \mu_i^2 k k' \cosh\left(\frac{\mu_i^2 k k'}{2}\right) 
- (\mu_i^2 k'^2 + 2) 
\sinh\left(\frac{\mu_i^2 k k'}{2}\right)\Bigg].
\end{align}
Notice that in Gogny forces the nucleon effective mass, besides being density dependent (through $\mu_n$ and $\mu_p$),
is momentum and temperature dependent.
The specific result for $f_q^{\gog}$ at $T=0$ is analytical and can be found in Refs.~\cite{arnau1,davesne2025}.

\begin{figure*}[!htb]
\centering
\includegraphics[scale=0.325]{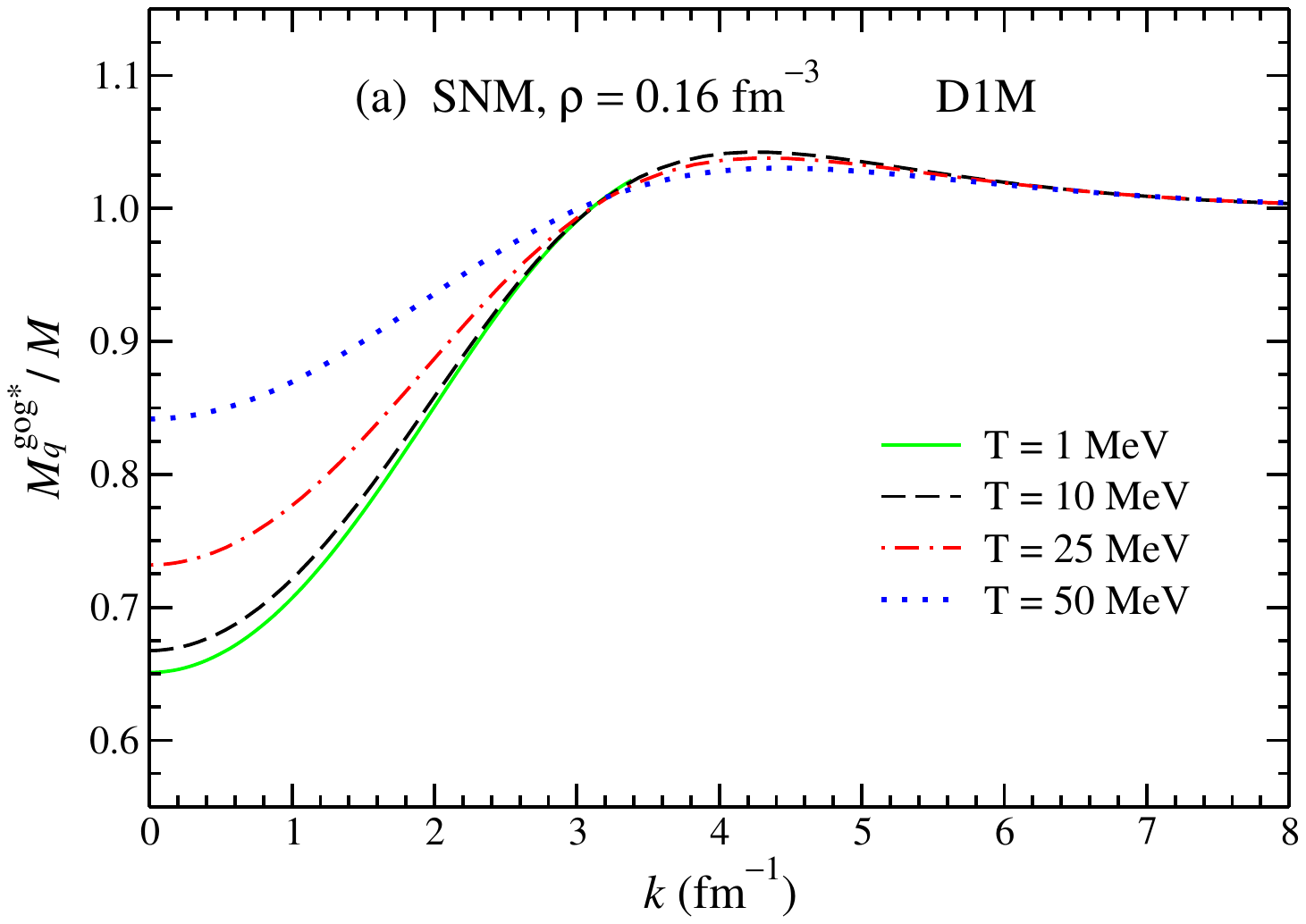}\hspace{-1.5cm}
\includegraphics[scale=0.325]{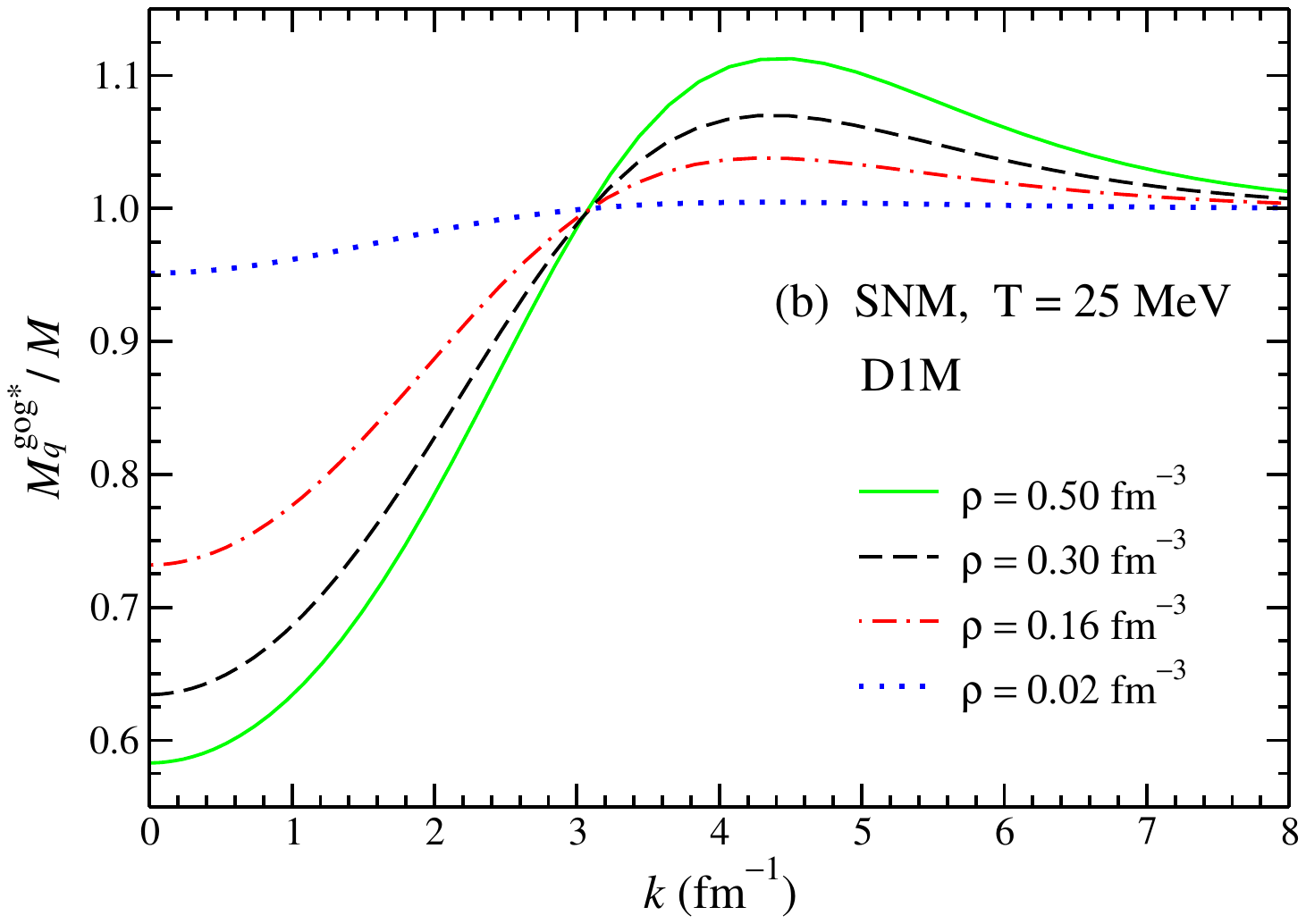}\vspace{-0.8cm}
\includegraphics[scale=0.325]{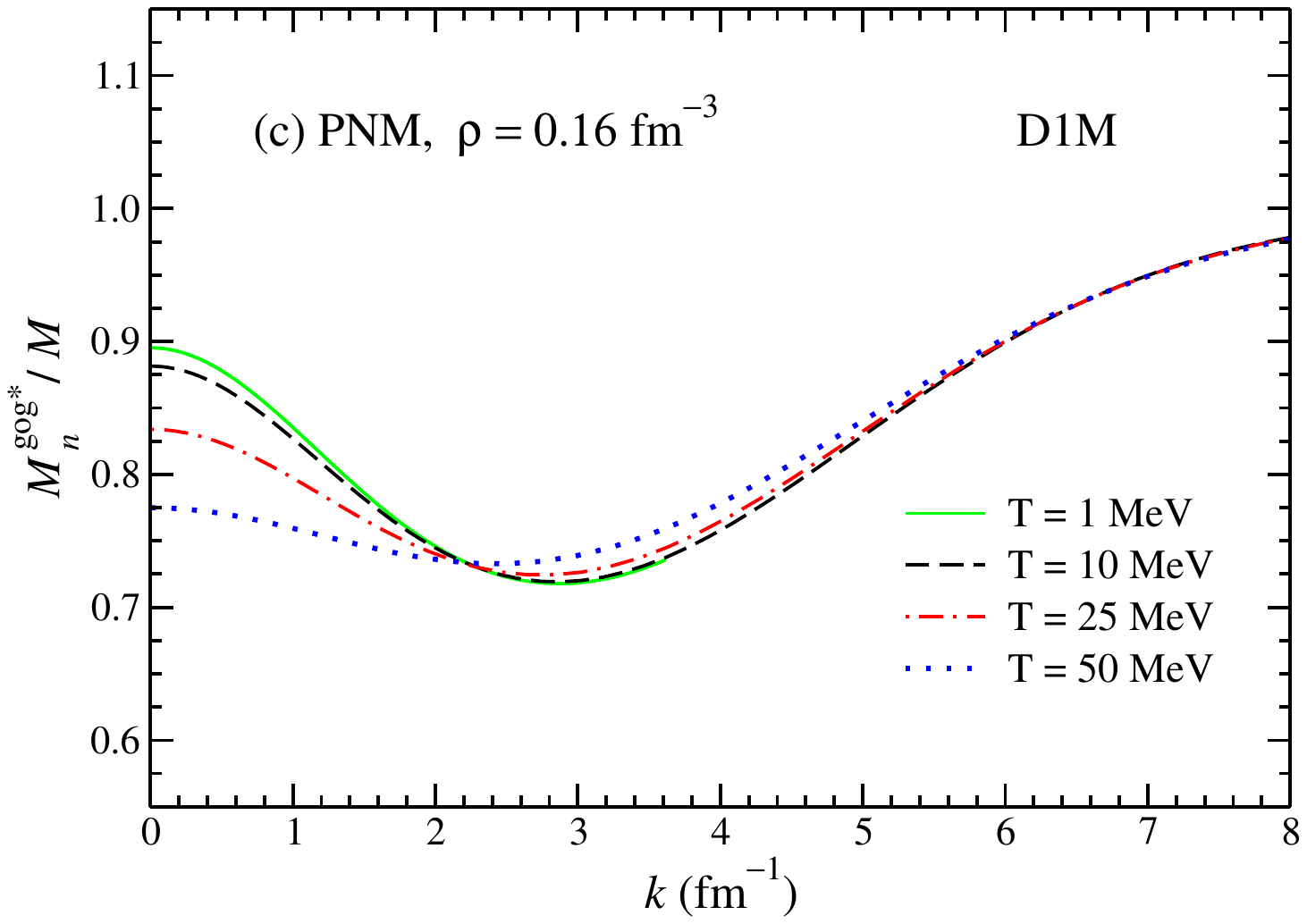}
\caption{Nucleon effective mass, as a function of momentum $k$, predicted by the Gogny D1M interaction for symmetric nuclear matter at $\rho=0.16$~fm$^{-3}$ and different temperatures (panel a), and at $T=25$~MeV and different values of $\rho$ (panel b). The effective mass of neutrons in pure neutron matter is also given (panel c).}
\label{fig:mgog}
\end{figure*}

We present in Fig.~\ref{fig:mgog} the $k$ dependence of $M_q^{\gog *}/M$ predicted by the Gogny D1M model in different systems, namely, symmetric nuclear matter (SNM) and pure neutron matter (PNM), at fixed values of density and temperature. For SNM, note that $M_q^{\gog *} = M_p^{\gog *} = M_n^{\gog *}$. For PNM, we show $M_q^{\gog *} = M_n^{\gog *}$. Our results for $M_q^{\gog *}$ at $T=1$ MeV in SNM and PNM are practically the same as those at $T=0$
shown for D1M in figure~4 of Ref.~\cite{arnau1}. Actually, for fixed density $\rho=0.16$~fm$^{-3}$, we see in Fig.~\ref{fig:mgog} that $M_q^{\gog *}$ changes little with temperature as long as $T \lesssim 10$ MeV. However, higher temperatures have a strong influence on the nucleon effective mass for momenta between $0$ and around~$3$~fm$^{-1}$.

We also provide, specifically at $T=0$ and for symmetric matter, the isovector effective mass for the Gogny model. Such a quantity is given by (see Appendix)
\begin{align}
&\frac{M}{M^{\gog *}_{\mathrm{vec}}} = 1 + \frac{2M}{\hbar^2} \frac{1}{\sqrt{\pi}} \sum_{i=1}^{2} \mu_i^2 \left[ 2M_i + H_i - B_i - \frac{W_i}{2} \right]
\nonumber\\
&\times \left[ \frac{2}{x^3} - \frac{1}{x} - \left( \frac{2}{x^3} + \frac{1}{x} \right) e^{-x^2} \right]_{x=\mu_i k_F}
\nonumber\\
&+ \frac{2M}{3\hbar^2} \frac{1}{\sqrt{\pi}} \sum_{i=1}^{2} \mu_i^2 \Bigg[ M_i + \frac{H_i}{2} - B_i - \frac{W_i}{2} \Bigg]
\nonumber\\
&\times \left[ \frac{1}{x} - \left( x + \frac{1}{x} \right) e^{-x^2} \right]_{x=\mu_i k_F} ,
\label{eq:mgogvec}
\end{align}
where $k_F$ is the Fermi momentum in SNM. This is one of the bulk parameters used to generate the Skyrme-versions of the Gogny parametrizations, as discussed in the next section.

\subsection{The Skyrme-version model}

In order to better identify the effect of the finite range interaction present in the aforementioned Gogny model, we construct here a zero-range Skyrme version of the respective Gogny parametrization. The equation of state of this zero-range model can be derived from the energy density. In the case of asymmetric matter at finite temperature, such a quantity is given by
\begin{align}
\mathcal{H}^{\sky}(\rho,\beta,T) = \sum_{q=n,p} \mathcal{H}^{\sky *}_{\kin,q}(\rho,\beta,T) + \mathcal{H}^{\sky}_{\pot}(\rho,\beta),
\end{align}
where, as before, $q=n,p$ is the isospin index. The kinetic part of the model reads
\begin{align}
\mathcal{H}^{\sky *}_{\kin,q} &= 2\int\!\!\frac{d\bf{k}}{(2\pi)^3} \frac{\hbar^2k^2}{2M^{\sky *}_q}F_{Dq}(k,T)
\label{eq:edenkinsky}
\end{align}
with the respective Fermi-Dirac distribution given by
\begin{align}
F_{Dq}(k,T) = \left\{1+e^{\left[\hbar^2k^2/(2M^{\sky *}_q) - \mu^*_{\kin,q}\right]/T}\right\}^{-1}.
\end{align}
Notice that for the Skyrme model we employ the symbol~$^*$ in the thermodynamical quantities related to the kinetic energy density, such as $\mathcal{H}^{\sky *}_{\kin,q}$ or $\mu^*_{\kin q}$, to make clear that they are functions of $M_q^{\sky *}$, and not $M$ (nucleon rest mass), i.e., all these quantities are affected by the in-medium dependence of the nucleon the effective mass. This notation is the same used in Refs.~\cite{dutra2023finite,lourenco2024}. The potential part of the energy density is~\cite{skyrme1,skyrme2,stone2007,dutra2012}
\begin{align}
\mathcal{H}^{\sky}_{\pot}(\rho,\beta) &= \frac{1}{8}t_0\rho^2[2(x_0+2)-(2x_0+1)H_2]
\nonumber\\
&+ \frac{1}{48}t_{3}\rho^{\alpha+2}[2(x_{3}+2)-(2x_{3}+1)H_2]\, ,
\label{eq:edenpotsky}
\end{align}
with
\begin{align}
H_2(\beta) = \frac{1}{2}[(1-\beta)^2 + (1+\beta)^2].
\end{align}
The asymmetry parameter is
\begin{align}
\beta= \frac{\rho_n-\rho_p}{\rho} = 1-2y \,,
\end{align}
with $y=\rho_p/\rho$ the proton fraction. The in-medium nucleon effective mass is expressed as~\cite{dutra2012}
\begin{eqnarray}
M_q^{\sky *}(\rho,\beta)=M\left\{1+\frac{M\rho}{4\hbar^2}[a + b(1+\tau_{3q}\beta)]\right\}^{-1}
\label{eq:skyefmass}
\end{eqnarray}
where
\begin{eqnarray}
a&=&t_1(x_1+2)+t_2(x_2+2)\mbox{,} \label{eq:a} \nonumber \\
b&=&\frac{1}{2}\left[t_2(2x_2+1)-t_1(2x_1+1)\right],
\label{eq:ab}
\end{eqnarray}
with $\tau_{3n}=1$ (neutrons) and $\tau_{3p}=-1$ (protons).

By adopting here the canonical ensemble, the thermodynamical potential that describes the system is the Helmholtz free energy density $\mathcal{F}^{\sky} = \mathcal{F}^{\sky *}_{\kin} + \mathcal{F}_{\pot}^\sky$. Pressure, entropy density, and chemical potentials are obtained from $\mathcal{F}^{\sky}$ and, since the effective mass of the Skyrme model is a temperature-independent quantity, it is possible to write $\mathcal{F}^{\sky}_{\pot} = \mathcal{H}^{\sky}_{\pot}$, according to the findings of Refs.~\cite{dutra2023finite,lourenco2024}. This feature leads to the following expression for the pressure,
\begin{align}
P^{\sky}(\rho,\beta,T) &= \rho^2 \frac{\partial (\mathcal{F}^{\sky}/\rho)}{\partial \rho}\Bigg|_{T,\beta}
\nonumber\\
&= \sum_{q=n,p} \left(P^{\sky *}_{\kin,q} + P^{\sky}_{\corr,q} \right)+P^{\sky}_{\pot},
\label{eq:psky}
\end{align}
where~\cite{lourenco2024}
\begin{align}
&P^{\sky *}_{\kin,q} = \frac{2}{3}\mathcal{H}^{\sky *}_{\kin,q},
\\
&P^{\sky}_{\corr,q} = - \frac{3}{2}\rho\frac{P^{\sky *}_{\kin,q}}{M^{\sky *}_q}\frac{\partial M^{\sky *}_q}{\partial\rho}\Bigg| _{T,\beta},
\\
&P^{\sky}_{\pot} = \frac{1}{8}t_0\rho^2[2(x_0+2)-(2x_0+1)H_2]
\nonumber\\
&+ \frac{1}{48}t_{3}(\alpha+1)\rho^{\alpha+2}[2(x_{3}+2)-(2x_{3}+1)H_2].
\end{align}
The nucleon entropy density reads
\begin{equation}
{\cal S}^{\sky}_q = - 2\int\!\!\frac{d\bf{k}}{(2\pi)^3} [F_{Dq}\mbox{ ln} F_{Dq} + (1-F_{Dq})\mbox{ln}(1-F_{Dq})],
\label{eq:entdenfg}
\end{equation}
and the expression for the chemical potentials is given by~\cite{lourenco2024}
\begin{align}
\mu_q^{\sky}(\rho,\beta, T) &= \mu^*_{\kin,q} + \mu_{\corr,q}^{\sky} + \mu_{\pot,q}^{\sky}
\label{eq:muqsky}
\end{align}
with
\begin{align}
\mu_{\corr,q}^{\sky} &= - \frac{3}{2}\frac{P^{\sky *}_{\kin,n}}{M^{\sky *}_n}\frac{\partial M^{\sky *}_n}{\partial \rho_q}\Bigg| _{T,n_{\bar{q}}}
\nonumber\\
&- \frac{3}{2}\frac{P^{\sky *}_{\kin,p}}{M^{\sky *}_p}\frac{\partial M^{\sky *}_p}{\partial \rho_q}\Bigg| _{T,n_{\bar{q}}},
\nonumber\\
\mu_{\pot,q}^{\sky} &= \frac{t_0}{4}\rho\left\{2(x_0+2)-(2x_0+1)[H_2\pm(1\mp\beta)\beta]\right\}
\nonumber\\
&+\frac{(\alpha+2)}{48}t_3\rho^{\alpha+1}\Bigg\{2(x_3+2) - (2x_3+1)
\nonumber\\
&\times\left[H_2\pm\frac{2(1\mp\beta)\beta}{\alpha+ 2}\right]\Bigg\}\, ,
\end{align}
with upper (lower) signs for neutrons (protons) and the same previous convention that, for a given particle isospin-index $q$, ${\bar{q}}$ is the other isospin-index.
The label ``corr'' in $P^{\sky}_{\corr,q}$ and $\mu_{\corr,q}^{\sky}$ indicates that they are the correction terms that arise from the variation of the Skyrme effective mass $M_q^{\sky *}$ with density. From Eq.~\eqref{eq:muqsky}, and by applying the relation $\hbar^2k^2/(2M^{\sky *}_q) - \mu^*_{\kin q} = \hbar^2k^2/(2M) + u^{\sky}_q - \mu_q^{\sky}$, it is possible to define the nucleon single-particle potential of the Skyrme model, in this case given by $u^{\sky}_q=k^2\rho[a + b(1+\tau_{3q}\beta)]/8+ \mu_{\corr,q}^{\sky} + \mu_{\pot,q}^{\sky}$. Notice that $u^{\sky}_q$ presents a quadratic momentum dependence, differently from the single-particle potential of the Gogny model.

For computing $\mu^*_{\kin,q}$ we proceed as follows. The nucleon density $\rho_q$, given by Eq.~(\ref{eq:dens}) with the Fermi--Dirac distribution $F_{Dq}(k,T)$ for the Skyrme model, namely,
\begin{align}
\rho_q = 2\int\!\!\frac{d\bf{k}}{(2\pi)^3} \left\{1+e^{\left[\hbar^2k^2/(2M^{\sky *}_q) - \mu^*_{\kin,q}\right]/T}\right\}^{-1},
\end{align}
provides the relationship between $\rho_q$ and $\mu^*_{\kin,q}$. For a given temperature~$T$, and for a particular value of~$\rho_q$, we invert the above equation in order to find, numerically, the corresponding value of $\mu^*_{\kin,q}$.

In order to construct a Skyrme parametrization associated with the Gogny one, one needs to determine the values of seven free parameters, namely, $t_0$, $t_3$, $x_0$, $x_3$, $\alpha$, $a$ and $b$. For this purpose, we impose the Skyrme-version to present the same nuclear empirical properties of the respective Gogny model. In particular, we choose to fix saturation density, binding energy, incompressibility, isoscalar effective mass, isovector effective mass, symmetry energy, and symmetry energy slope. All quantities are evaluated at zero temperature and at the saturation density. For the Skyrme model, such bulk parameters are given, in the isoscalar sector, by $\rho_0$, $B^{\sky}_0=\mathcal{H}^{\sky}(\rho_0,0,0)/\rho_0$, $K^{\sky}_0=9[\partial P^{\sky}(\rho,0,0)/\partial\rho]_{\rho_0}$, and $M^{\sky *}_p(\rho_0,0)=M^{\sky *}_n(\rho_0,0)$. The quantities related to the isovector sector are the following: $M^{\sky *}_{\mathrm{vec}}(\rho_0)=M^{\sky *}_p(\rho_0,1)$~\cite{dutra2012,baoanli2018,vangiai2006} (also see the Appendix), $E^{\sky}_{\sym}\equiv E^{\sky}_{\sym}(\rho_0)=[\mathcal{H}^{\sky}(\rho_0,1,0)-\mathcal{H}^{\sky}(\rho_0,0,0)]/\rho_0$, $L^{\sky}_{\sym}\equiv L^{\sky}_{\sym}(\rho_0)=3\rho_0[\partial E^{\sky}_{\sym}(\rho)/\partial\rho]_{\rho_0}$. Notice that here, as for the Gogny model, we take the full definition of the symmetry energy, i.e., the difference between the energy per particle in pure neutron matter and symmetric nuclear matter.

It is worth mentioning that we have used the procedure described in Ref.~\cite{lourenco2024} for the derivation of the aforementioned equations of state since the Skyrme model presents in-medium effective nucleon mass and a momentum-independent potential at the mean-field level. In these cases, pressure and chemical potentials are impacted by the emergence of correction terms induced by the density dependence of the effective mass. As a remark, notice that the medium dependent effective mass in the Skyrme model just appears because of the particular momentum dependence (quadratic) of the Skyrme interaction that allows to shift part of the effective interaction to the kinetic energy term.

Concerning the numerical implementation of the Skyrme-version model, we treat its Fermi integrals by using the analytical approach proposed in Ref.~\cite{jel}. Such a method was shown to be a fast and accurate approximation for Skyrme and relativistic mean-field models, and not only for the free Fermi gas, as the reader can verify in the study performed in Refs.~\cite{dutra2023finite,lourenco2024}.

\section{Warm infinite nuclear matter}
\label{sec:results}

\begin{figure}[!htb]
\centering
\includegraphics[scale=0.53]{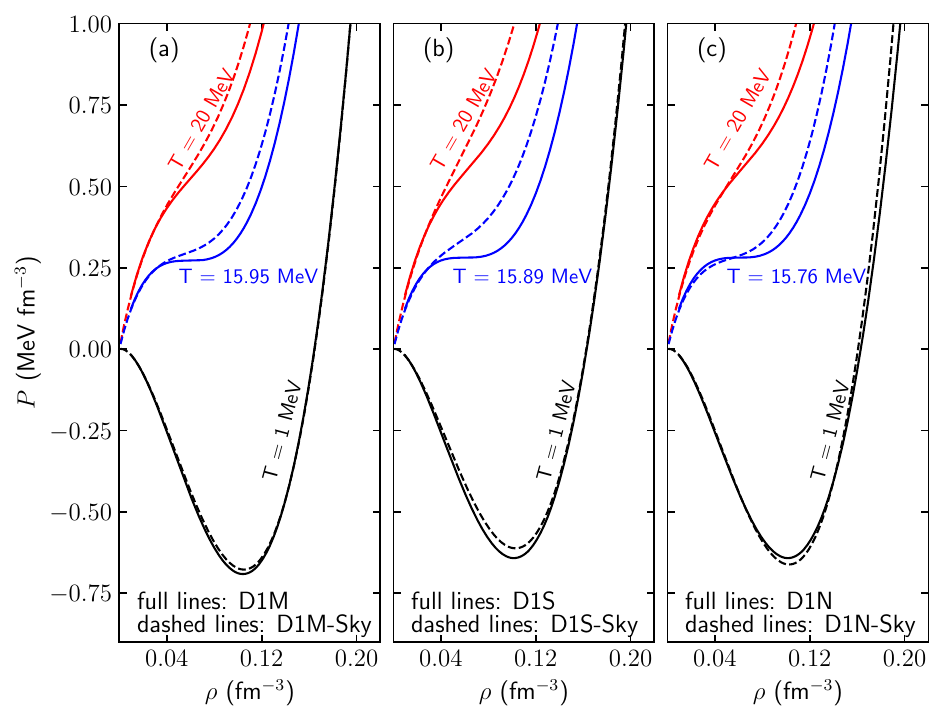}
\caption{Pressure as a function of $\rho$ of symmetric matter ($\beta=0$) for distinct Gogny parametrizations (D1M \cite{PRL102Goriely2009}, D1S \cite{Berger1991}, and D1N \cite{PLB668Chappert2008}) compared with their respective Skyrme-versions, all of them at different temperatures. Full (dashed) lines: Gogny (Skyrme-version) model. The temperature related to the isotherms marked in blue color corresponds to the critical temperature of the Gogny interactions.}
\label{fig:press_SM}
\end{figure}

As a first analysis, we present in Fig.~\ref{fig:press_SM} the density dependence of the pressure for some parametrizations of the Gogny and Skyrme-version models for symmetric nuclear matter at finite temperature.
From the figure, one observes a systematic behavior concerning the isotherms. All Skyrme-version parametrizations present more significant differences in comparison with the respective Gogny ones as the temperature increases. We also obtain the critical parameters related to the liquid-gas phase transition for both models. For symmetric nuclear matter, they are calculated through the following conditions
\begin{align}
\frac{\partial P(\rho,0,T)}{\partial\rho}\Big|_{\rho=\rho_c,T=T_c} = 0,
\label{eq:cond1}
\\
\frac{\partial^2 P(\rho,0,T)}{\partial\rho^2}\Big|_{\rho=\rho_c,T=T_c} = 0,
\label{eq:cond2}
\end{align}
that allow us to extract the density~($\rho_c$) and temperature~($T_c$) at the critical point. These quantities are used to calculate the critical pressure defined as $P_c=P(\rho_c,0,T_c)$. The associated critical isotherm in the pressure versus density space ($T=T_c$) is characterized by the full agreement with the mechanical stability condition, given by
\begin{align}
\frac{\partial P(\rho,0,T)}{\partial\rho}\Big|_{T} > 0.
\label{eq:mecstab}
\end{align}
In Table~\ref{tab:snm} we furnish $\rho_c$, $P_c$, and $T_c$ along with the nuclear bulk parameters. In the case of symmetric matter, only $\rho_0$, $B_0$, $K_0$ and $m^*_0=M^*(\rho_0,0)/M$ are relevant in order to generate the Skyrme
versions.
\begin{table*}[!htb]
\tabcolsep=0.19cm
\centering
\caption{Critical parameters (SNM case) for the Gogny parametrizations and Skyrme versions used here. The isoscalar effective mass ratio $m^*_0$ is given in Refs.~\cite{dutra2021,arnau1}, and the remaining nuclear bulk properties can be found in Ref.~\cite{claudia2017}, except for the isovector effective mass ratio $m^*_{\mathrm{v}0}=M^*_{\mathrm{vec}}(\rho_0)/M$, that is presented in this work. All bulk parameters are the same for the respective Skyrme versions.}
\begin{tabular}{l|c|c|c|c|c|c|c|c|c|c|c|c}
\hline\hline
Models & $T_c$ & $\rho_c$ & $P_c$ & $\frac{\rho_c}{\rho_0}$ & $\frac{P_c}{\rho_c T_c}$ & $B_0$ & $\rho_0$ & $K_0$ & $m^*_0$ & $E_\sym$ & $L_\sym$ & $m^*_{\mathrm{v}0}$ \\
      & (MeV) & (fm$^{-3}$) & (MeV/fm$^3$) &   &   & (MeV) & (fm$^{-3}$) & (MeV) &  & (MeV) & (MeV) & \\
\hline
D1S     & 15.89 & 0.060 & 0.281 & 0.368 & 0.295 & $-$16.01 & 0.163 & 202.88 & 0.697 & 31.95 & 22.28 & 0.575 \\
D1S-Sky & 13.64 & 0.054 & 0.194  & 0.329 & 0.265 &        &       &        &       &       &       &       \\
D1M     & 15.95 & 0.058 & 0.272 & 0.352 & 0.294 & $-$16.03 & 0.165 & 224.98 & 0.746 & 29.73 & 24.67 & 0.718 \\
D1M-Sky & 14.77 & 0.057 & 0.224 & 0.331 & 0.278 &        &       &        &       &       &       &       \\
D1N     & 15.76 & 0.056 & 0.261 & 0.348 & 0.296 & $-$15.96 & 0.161 & 225.65 & 0.747 & 30.14 & 31.95 & 0.761 \\
D1N-Sky & 14.70 & 0.053 & 0.219 & 0.331 & 0.279 &        &       &        &       &       &       &       \\
\hline\hline
\end{tabular}
\label{tab:snm}
\end{table*}
It is worth noticing the systematic reduction of all critical parameters of the Skyrme-versions in comparison with the respective original Gogny parametrizations. For the sake of completeness, we also furnish the parameters of the corresponding Skyrme-versions parametrizations. These numbers are shown in Table.~\ref{tab:parsky}.
\begin{table}[!htb]
\tabcolsep=0.2cm
\centering
\caption{Parameter sets of the Skyrme-version model used in this work.}
\begin{tabular}{l|r|c|c}
\hline\hline
Parameter set                  & D1S-Sky     & D1M-Sky   & D1N-Sky   \\
\hline
$t_0$ (MeV$\cdot$fm$^3$)             & 16127.35    & $-$2550.033 & $-$2443.908 \\
$t_3$ (MeV$\cdot$fm$^{3\alpha + 3}$) &$-$98508.10    & 14563.570 & 14067.18  \\
$x_0$                          & 1.012974    & 0.7154947 & 0.6731205 \\
$x_3$                          & 0.9772651   & 1.033744  & 1.024757  \\
$a$ (MeV$\cdot$fm$^5$)               & 752.7825    & 395.035   & 324.2818  \\
$b$ (MeV$\cdot$fm$^5$)               & $-$310.3950   & $-$52.74730 & 24.66134  \\
$\alpha$                       & $-$0.01642706 & 0.1652792 & 0.1800967 \\
\hline\hline
\end{tabular}
\label{tab:parsky}
\end{table}

\begin{figure*}[!htb]
\centering
\includegraphics[scale=0.55]{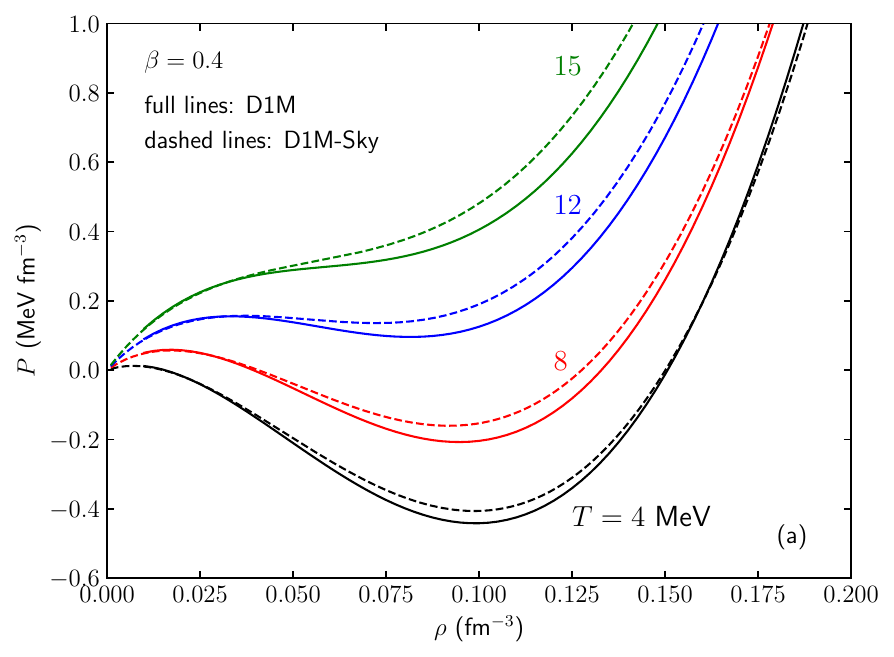}
\includegraphics[scale=0.55]{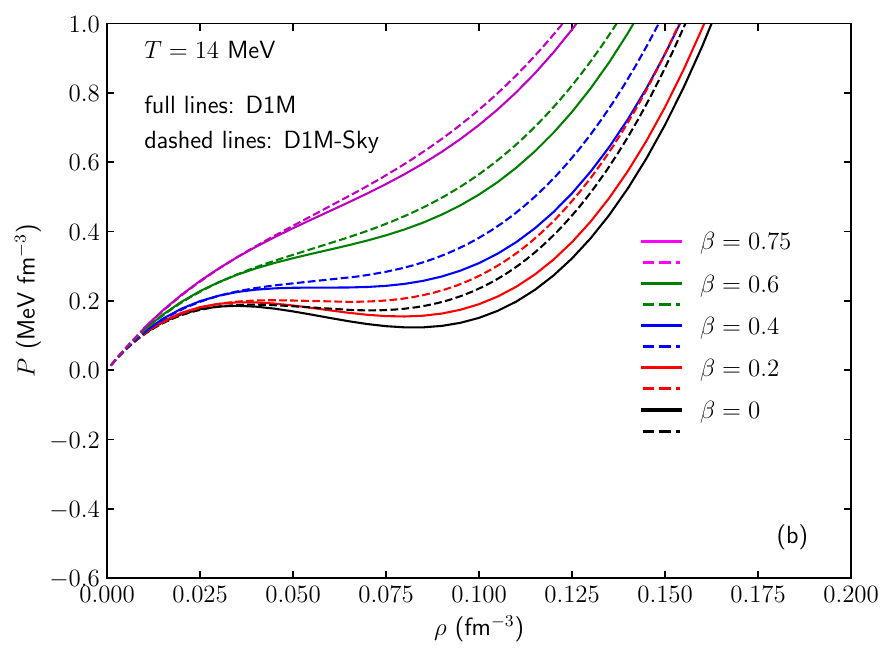}
\caption{Pressure as a function of density $\rho$ of asymmetric matter for the Gogny model D1M (full curve) and for its Skyrme-version model D1M-Sky (dashed curve). Results in panel (a) are for an asymmetry parameter
$\beta=0.4$ and different temperatures. Results in panel (b) are for a temperature $T = 14$~MeV and different asymmetry parameters.}
\label{fig:press_ANM}
\end{figure*}

In Fig.~\ref{fig:press_ANM}, we present our results for asymmetric nuclear matter (ANM). The pressure is displayed as a function of density for different values
of temperature (panel a), and asymmetry parameters (panel b).
As we see, the same feature exhibited in Fig.~\ref{fig:press_SM} is also observed in Fig.~\ref{fig:press_ANM}{\color{blue}a}, i.e., strong deviations between the Gogny and Skyrme-version models for higher temperatures. For a fixed temperature, on the other hand, the effect is inversely proportional to the asymmetry parameter, namely, the smaller $\beta$ the greater the difference between the respective curves, at least for a sufficiently high temperature. In Fig.~\ref{fig:press_ANM}{\color{blue}b} we compare the models for a fixed temperature, $T=14$~MeV in this case, and different values of the asymmetry parameter $\beta$. Notice that smaller deviations are verified for higher values of $\beta$.

\begin{figure}[!htb]
\centering
\includegraphics[scale=0.32]{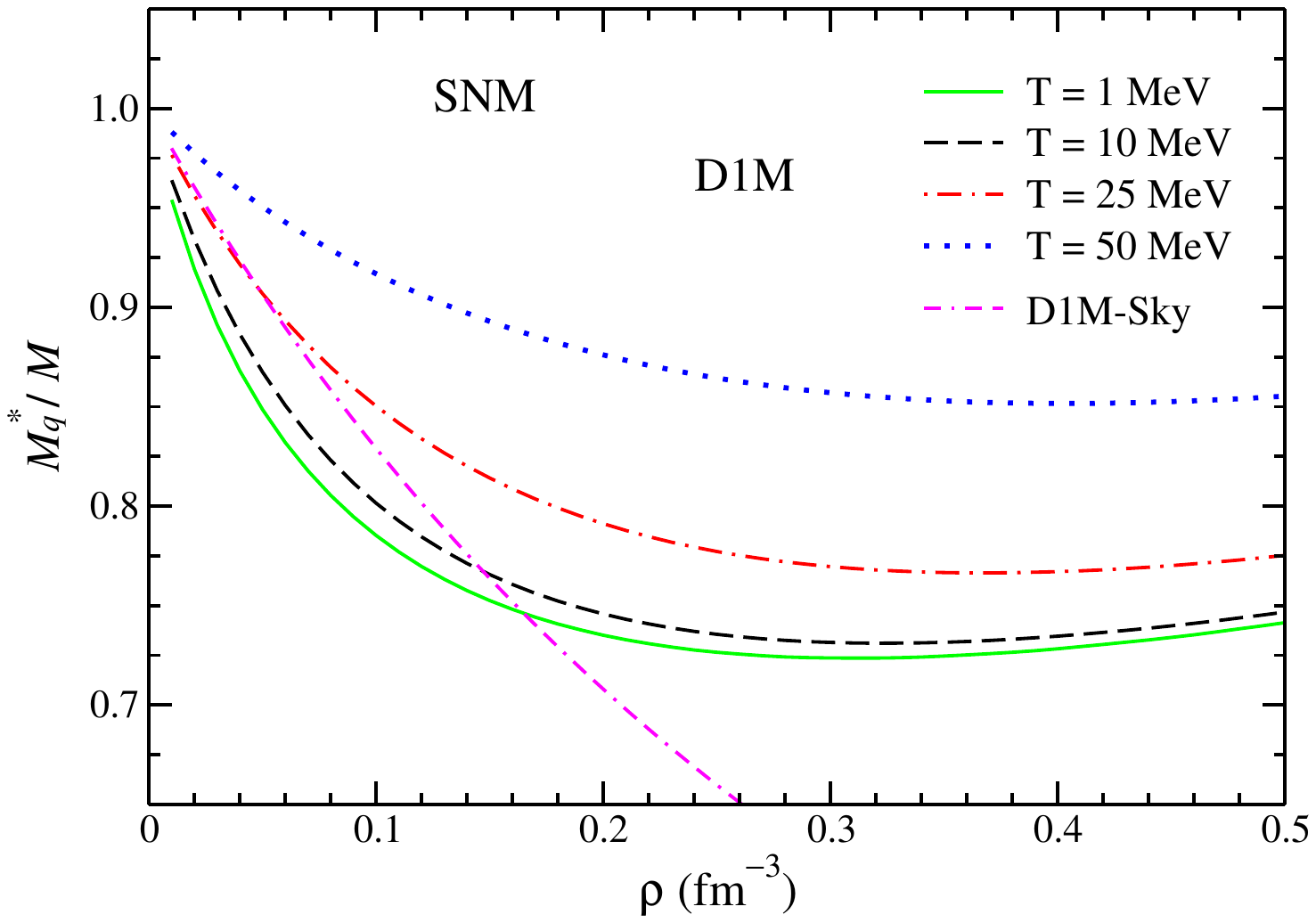}
\caption{Nucleon effective mass, as a function of density, predicted by the Gogny D1M interaction for SNM at different temperatures. The respective Skyrme version (D1M-Sky) is also shown.}
\label{fig:mstar_rho}
\end{figure}

For completeness, in Fig.~\ref{fig:mstar_rho} we plot for several temperature values the isoscalar effective mass of the Gogny D1M interaction in SNM
as a function of density, with the momentum dependence evaluated at the cold Fermi momentum, namely, $k= k_F= (3\pi^2 \rho/2)^{1/3}$. We also show the respective curve for the Skyrme version of such an interaction. 
Notice that in the Skyrme model, the effective mass depends only on the density and, therefore, it is not affected by temperature, see Eq.~\eqref{eq:skyefmass}. The situation is different for Gogny forces, where the effective mass is affected by temperature due to its momentum dependence, as can be seen in Fig.~\ref{fig:mgog}. Furthermore, from Fig.~\ref{fig:mstar_rho} one can see that $M_q^{\gog *}/M$ increases with temperature as a result of the more diffuse momentum distribution. This pattern is compatible with the increasing trend with the temperature of $M_q^{\gog *}/M$ as a function of $k$, see Fig.~\ref{fig:mgog}{\color{blue}a}. The crossing between the effective mass computed with Gogny (at $T=1$~MeV) and the Skyrme-equivalent forces occurs at the saturation density as expected, as long as we have imposed such a constraint in the fitting of the Skyrme-equivalent parameters. Such features indicate that the Gogny effective mass is much richer than the one of Skyrme because of its $k$-dependence, which has a real impact on the temperature behavior of the isoscalar effective mass, but also on its dependence on $k_F$ at $T=0$~MeV, which is very different from the one predicted by the Skyrme model.

\subsection{Liquid-gas phase transition}

With regard to the liquid-gas phase transition in ANM, it is known that such unbalanced isospin two-fluid (binary) system offers a richer phase diagram in comparison with the one obtained in the symmetric case. One clear difference comes from the Gibbs criteria for phase coexistence, generalized from
\begin{align}
P(\rho_a,0,T) &= P(\rho_b,0,T),
\label{eq:gibbssm1}
\\
\mu(\rho_a,0,T) &= \mu(\rho_b,0,T),
\label{eq:gibbssm2}
\end{align}
in SNM to
\begin{align}
P(\rho_a,\beta_a,T) &= P(\rho_b,\beta_b,T),
\label{eq:gcpress}
\\
\mu_p(\rho_a,\beta_a,T) &= \mu_p(\rho_b,\beta_b,T),
\label{eq:gcmup}
\\
\mu_n(\rho_a,\beta_a,T) &= \mu_n(\rho_b,\beta_b,T),
\label{eq:gcmun}
\end{align}
for ANM, with the indexes $a$ and $b$ referring to the different phases of the system (gas and liquid).

In the SNM case, the boundaries of the thermodynamical phases are defined by $\rho_a$ and $\rho_b$, both calculated from the solution of Eqs.~\eqref{eq:gibbssm1}-\eqref{eq:gibbssm2} at a given temperature~$T$. This set of equations generates $\rho_a\neq\rho_b$ in the range of $T<T_c$, and specifically at $T=T_c$ one has $\rho_a=\rho_b=\rho_c$. At this point, the conditions~\eqref{eq:cond1}-\eqref{eq:cond2}
are fully satisfied since the resulting isotherm presents an inflection point. Therefore, the Gibbs criteria offer an alternative procedure in order to obtain the critical parameters of the system. 

\begin{figure}[t]
\centering
\includegraphics[scale=0.32]{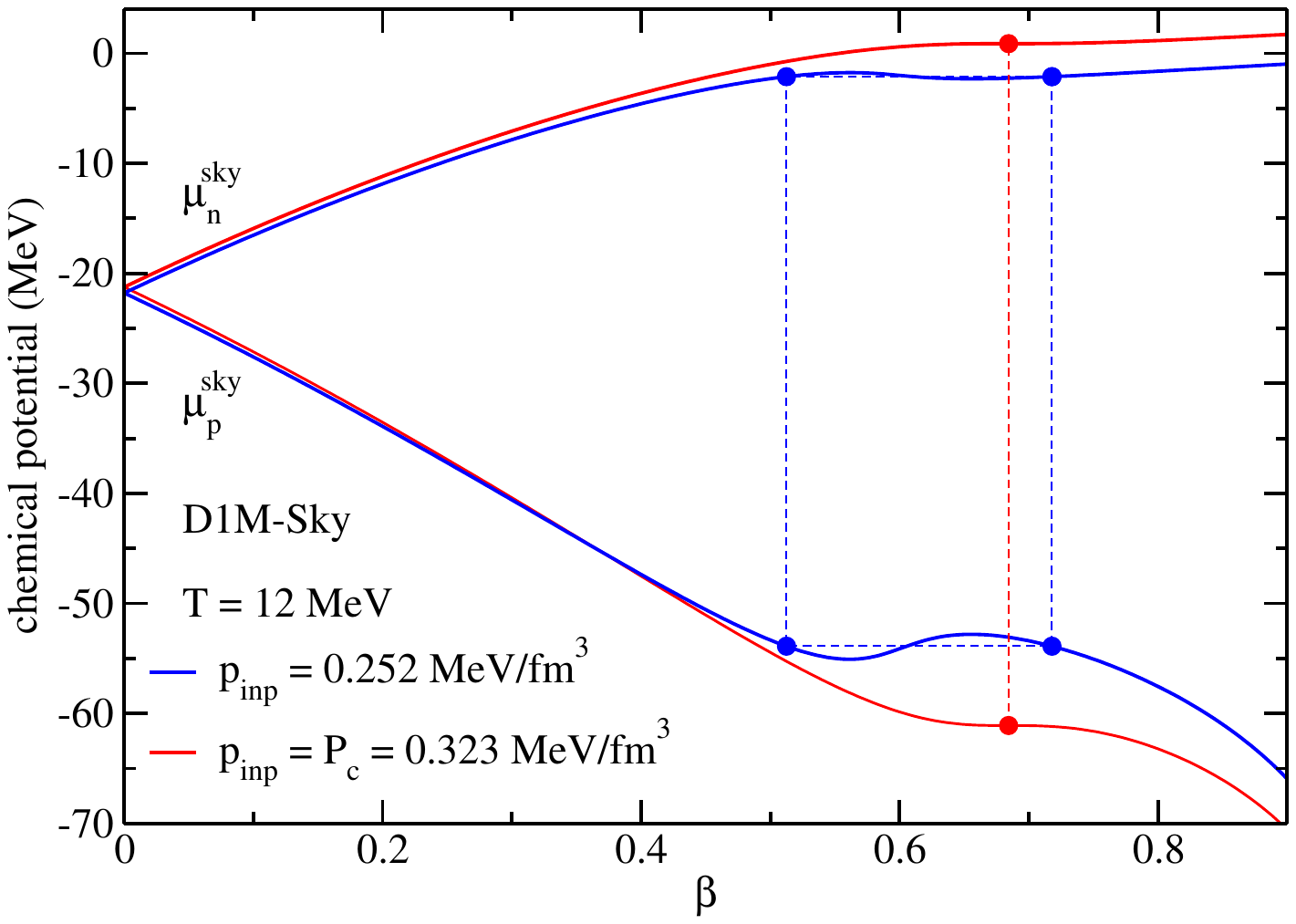}
\caption{Proton and neutron chemical potentials for the \mbox{D1M-Sky} parametrization at $T=12$~MeV. Results for $p_\inp=0.252$~MeV/fm$^3$ (blue curves) and $p_\inp=0.323$~MeV/fm$^3$ (red curves). Rectangle and vertical line: geometrical construction for the binodal section (see text).}
\label{fig:chpot}
\end{figure}

For binary systems (ANM), Eq.~\eqref{eq:gcpress} is actually splitted into two equations: $P(\rho_a,\beta_a,T) = p_\inp$ and $P(\rho_b,\beta_b,T) = p_\inp$. Then, the new set to be solved encompasses four equations used to obtain $\rho_a$, $\beta_a$, $\rho_b$, and $\beta_b$. It is possible to find these solutions from the so-called geometrical construction in the chemical potential space: at a particular value of temperature and input pressure ($p_\inp$), one looks for the points in $\mu_p$ and $\mu_n$, as functions of $\beta$,
that simultaneously satisfy conditions \eqref{eq:gcmup}-\eqref{eq:gcmun}. Technically, for each value
of $\beta$ we solve equation $P(\rho,\beta,T)-p_\inp=0$, with given $T$ and $p_\inp$, to find the related density~(densities). Therefore, $\mu_p(\rho,\beta,T)$ and $\mu_n(\rho,\beta,T)$ are evaluated as functions of $\beta$. We show in Fig.~\ref{fig:chpot} an example of this construction for the D1M-Sky parametrization.
Notice that for each pair $(T,p_\inp)$ there is only one possible rectangle corresponding to $\beta_a$~(gaseous-phase boundary at low density and high asymmetry) and $\beta_b$~(liquid-phase boundary at high density and low asymmetry) with respective densities~$\rho_a$,~$\rho_b$. Specifically for $T=12$~MeV and $p_\inp=0.252$~MeV/fm$^3$, we find $\beta_a=0.718$, $\beta_b=0.513$, $\rho_a=0.0364$~fm$^{-3}$, and $\rho_b=0.244$~fm$^{-3}$ (blue lines in Fig.~\ref{fig:chpot}).

With regard to the sections of the binodal surfaces under isothermal compression for the Skyrme-version models, we construct such phase diagrams in the pressure versus asymmetry parameter plot by using the aforementioned procedure, and by running the input pressure in the range of $p_\mi<p_\inp<p_\ma$ for a specific value of $T$. The results are presented in the next Figs.~\ref{fig:D1Mbin}-\ref{fig:D1Sbin}.
\begin{figure*}[!htb]
\centering
\includegraphics[scale=0.6]{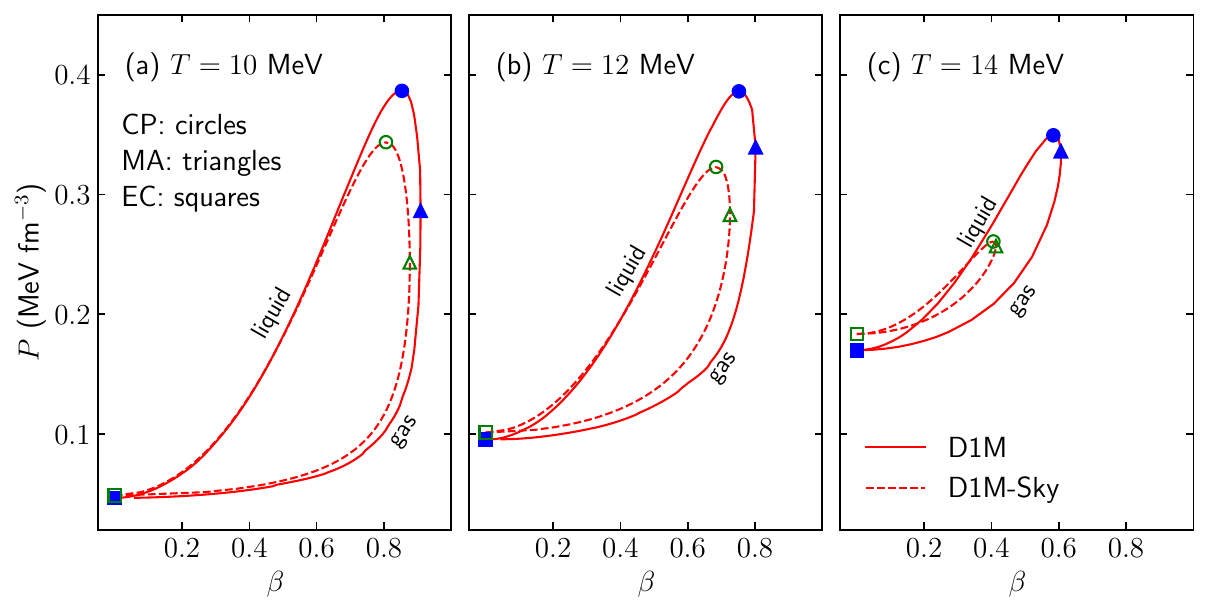}
\caption{Binodal sections at (a)~$T=10$~MeV, (b)~$T=12$~MeV, and (c)~$T=14$~MeV for D1M (solid lines) and D1M-Sky (dashed lines) parametrizations. The critical point (CP), the maximal isospin asymmetry (MA) point, and the equal concentration (EC) point are depicted by circles, triangles, and squares, respectively.}
\label{fig:D1Mbin}
\end{figure*}
\begin{figure*}[!htb]
\centering
\includegraphics[scale=0.6]{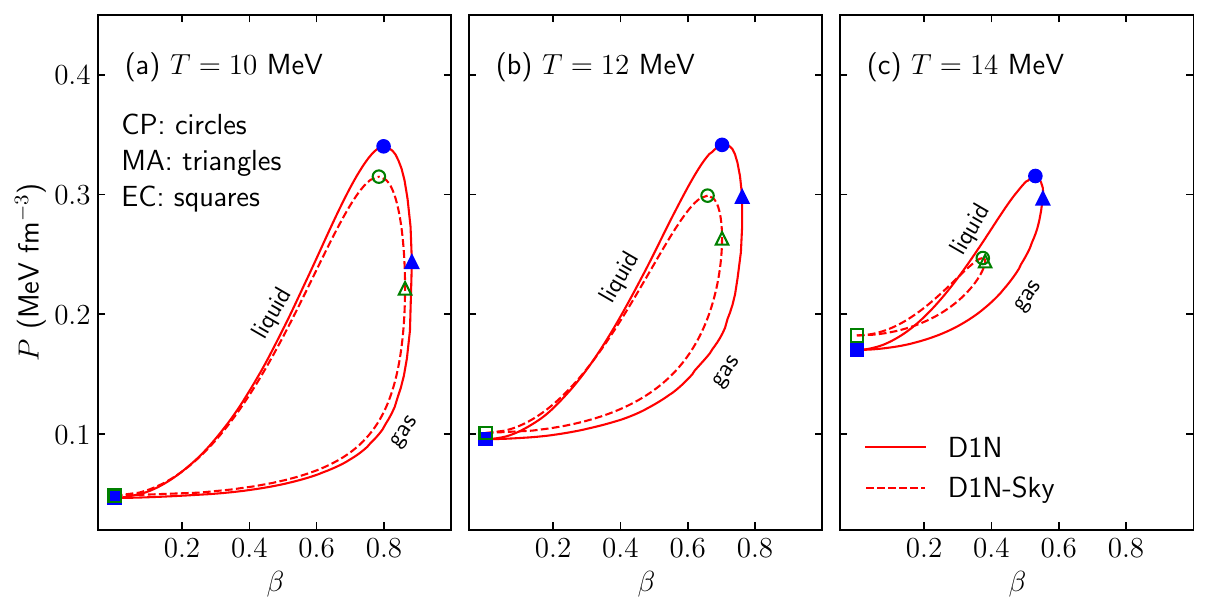}
\caption{Binodal sections at (a)~$T=10$~MeV, (b)~$T=12$~MeV, and (c)~$T=14$~MeV for D1N and D1N-Sky parametrizations.}
\label{fig:D1Nbin}
\end{figure*}
\begin{figure*}[!htb]
\centering
\includegraphics[scale=0.6]{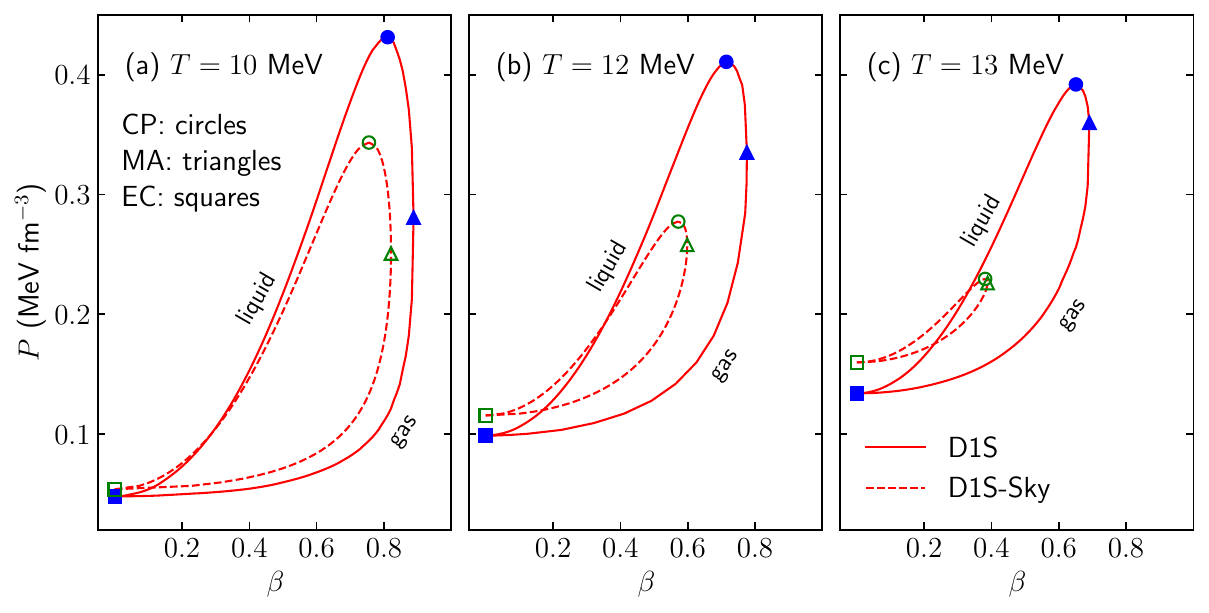}
\caption{Binodal sections at (a)~$T=10$~MeV, (b)~$T=12$~MeV, and (c)~$T=13$~MeV for D1S and D1S-Sky parametrizations.}
\label{fig:D1Sbin}
\end{figure*}

At this point we remark to the reader the existence of another method, different from the Gibbs conditions, that can also be applied for the description of the liquid-gas phase transition in asymmetric nuclear matter. It was proposed in Ref.~\cite{ducoin2006} (and used also in Ref.~\cite{typel2014}). The procedure is based on a suitable change of the thermodynamical potential that describes the system through a Legendre transform of the Helmholtz free energy and the pressure. By doing so, the new potential is then dependent on only one density and the phase coexistence is reduced to a one-component problem. Therefore, the usual Maxwell construction can be applied in this calculation. We have applied this method for computing the binodal sections of the Gogny models. We have found that in the case of the Gogny parametrizations this method turns out to be faster and allows us to obtain more accuracy than the direct solution of Eqs.~\eqref{eq:gcpress}-\eqref{eq:gcmun} that we have used with the Skyrme version models, as previously described.

From Figs.~\ref{fig:D1Mbin}-\ref{fig:D1Sbin}, we verify that $p_\mi$ and $p_\ma$ depend, in general, on both, temperature and the parametrization itself. The ($\beta_\mi,p_\mi$) point corresponds to $\beta_\mi=0$, i.e., the symmetric matter case, and because of that it is called point of equal concentration (EC), in which $p_\mi=p_\ec$. The pressure $p_\ma$, on the other hand, is related to the asymmetric matter critical point~(CP), namely, the point in which the binary system reaches the chemical stability boundary: $p_\ma=P_c$ for $\beta=\beta_c$. These two points, critical and EC ones, split the binodal section into a more asymmetric low-density branch (gaseous phase), and a less asymmetric high-density one (liquid phase).

It is possible to obtain $P_c$ from the geometrical construction in the chemical potential isobars by looking at the value of $p_\inp$ that converts the rectangle into a vertical line (red line in Fig.~\ref{fig:chpot}). The extremes of such a line indicate the inflection points of $\mu_p$ and $\mu_n$. At $p_\inp=P_c$, the proton (neutron) chemical potential always decreases (increases) as the asymmetry parameter increases. Such a feature is fully compatible with the chemical stability condition written as
\begin{align}
\frac{\partial \mu_p(\beta,P,T)}{\partial\beta}\Big|_{P,T} < 0,
\label{eq:chemstab1}
\end{align}
or
\begin{align}
\frac{\partial \mu_n(\beta,P,T)}{\partial\beta}\Big|_{P,T} > 0.
\label{eq:chemstab2}
\end{align}

A practical way of finding the critical pressure ($P_c$) and critical asymmetry parameter ($\beta_c$), at a particular temperature, is by simultaneously solving the set
\begin{align}
\frac{\partial \mu_p(y,P,T)}{\partial y}\Big|_{y=y_c,P=P_c,T} = 0,
\label{eq:condmu1}
\end{align}
\begin{align}
\frac{\partial^2 \mu_p(y,P,T)}{\partial y^2}\Big|_{y=y_c,P=P_c,T} = 0,
\label{eq:condmu2}
\end{align}
where the relationship between the critical proton fraction $y_c$ and $\beta_c$ is $\beta_c=1-2y_c$. However, since the chemical potentials of Gogny and Skyrme-version models are written in terms of $T$, $\beta$ (or equivalently $y$), and $\rho$ instead of $P$, it is suitable to invoke the following thermodynamical relation~(see~\cite{lee2003,dutra2012}, for instance),
\begin{align}
\left(\frac{\partial\mu_q}{\partial y}\right)_{P,T} &=
\left(\frac{\partial\mu_q}{\partial y}\right)_{\rho,T}
- \frac{(\partial P/\partial y)_{\rho,T}}{(\partial P/\partial \rho)_{y,T}}
\left(\frac{\partial\mu_q}{\partial \rho}\right)_{y,T}
\nonumber\\
&\equiv f(\rho,y,T),
\label{eq:dmudy}
\end{align}
and use it to also find
\begin{widetext}
\begin{align}
\left(\frac{\partial^2\mu_q}{\partial y^2}\right)_{P,T}& =
\left(\frac{\partial^2\mu_q}{\partial y^2}\right)_{\rho,T}
- \frac{(\partial^2 P/\partial y^2)_{\rho,T}}{(\partial P/\partial \rho)_{y,T}}
\left(\frac{\partial\mu_q}{\partial \rho}\right)_{y,T}
+ 2\left(\frac{\partial^2P}{\partial\rho\,\partial y}\right)\frac{(\partial P/\partial y)_{\rho,T}}{[(\partial P/\partial \rho)_{y,T}]^2}\left(\frac{\partial\mu_q}{\partial \rho}\right)_{y,T}
\nonumber\\
&- 2\frac{(\partial P/\partial y)_{\rho,T}}{(\partial P/\partial \rho)_{y,T}}\left(\frac{\partial^2\mu_q}{\partial\rho\,\partial y}\right)_T
- \left(\frac{\partial^2P}{\partial\rho^2}\right)_{y,T}\frac{[(\partial P/\partial y)_{\rho,T}]^2}{[(\partial P/\partial \rho)_{y,T}]^3}\left(\frac{\partial\mu_q}{\partial \rho}\right)_{y,T}
+ \left[\frac{(\partial P/\partial y)_{\rho,T}}{(\partial P/\partial \rho)_{y,T}}\right]^2\left(\frac{\partial^2\mu_q}{\partial\rho^2}\right)_{y,T}
\nonumber\\
&\equiv g(\rho,y,T).
\label{eq:d2mudy2}
\end{align}
\end{widetext}
Therefore, the conditions given in Eqs.~\eqref{eq:condmu1}-\eqref{eq:condmu2} can be replaced by the ones containing $\rho$, $y$, and $T$: $f(\rho_c,y_c,T) = 0$ and $g(\rho_c,y_c,T) = 0$. From this procedure, we find $\rho_c$ and $y_c$, at a given $T$, and since these quantities are obtained, the evaluation of the critical pressure is straightforward: $P_c=P(\rho_c,\beta_c,T)$.

Still concerning the results displayed in Figs.~\ref{fig:D1Mbin}-\ref{fig:D1Sbin}, we emphasize some other interesting features. The first one is related to the critical point. Notice that the values of the critical pressure and the critical asymmetry are systematically reduced in the Skyrme-like versions in comparison with the original Gogny forces. Such a finding, $P_c^\sky<P_c^\gog$, is also observed in the SNM case~($\beta=0$), as pointed out in Table~\ref{tab:snm}. It is also worth noting that, for each parametrization analyzed, the difference between $P_c^\gog$ and $P_c^\sky$ is more significant as the temperature increases. Similar patterns also occur with the point of maximal isospin asymmetry~(MA), also indicated in the figures. We see that the values of the asymmetry $\beta_\masy$ and of the pressure of the MA point are systematically smaller in the Skyrme-like models than in the Gogny models and that this discrepancy significantly increases with increasing temperature.

As a general trend shown in the figures, we observe a shrinkage of the binodal sections in the Skyrme-version models when temperature increases. As a consequence, the zero-range models tend to collapse to a single point, at $T=T_c$ of SNM, faster than the respective Gogny forces. Another important remark in our results is the fact that no limiting pressure~ ($P_{\mbox{\tiny lim}}$) is found in the Gogny models used here, i.e., there is not a specific value of pressure $P_{\mbox{\tiny lim}}$ above which Eqs.~\eqref{eq:gcmup}-\eqref{eq:gcmun} have no solution before reaching the critical pressure $P_c$. The occurrence of a limiting pressure $P_{\mbox{\tiny lim}} < P_c$ is a feature observed in the finite range model studied in Ref.~\cite{xu07}, namely, the isoscalar momentum-dependent interaction eMDYI. In this particular force, the isoscalar part of the single-nucleon potential is momentum-dependent but the isovector part is not. The existence of $P_{\mbox{\tiny lim}}$ implies that a critical pressure can not be reached in the binodal section, see for instance Fig.~4 of Ref.~\cite{xu07}. The resulting limitation in the binodal section for this interaction is due to the fact that the nucleon chemical potentials present an asynchronous variation with pressure, i.e., $\mu_n$ increases more rapidly with pressure than $\mu_p$, as the reader can verify in Fig.~3b of Ref.~\cite{xu07}. This feature is not verified in the parametrizations of the Gogny models analyzed in this work.

\subsection{Stability in asymmetric nuclear matter}

\begin{figure*}[!htb]
\centering
\includegraphics[scale=0.7]{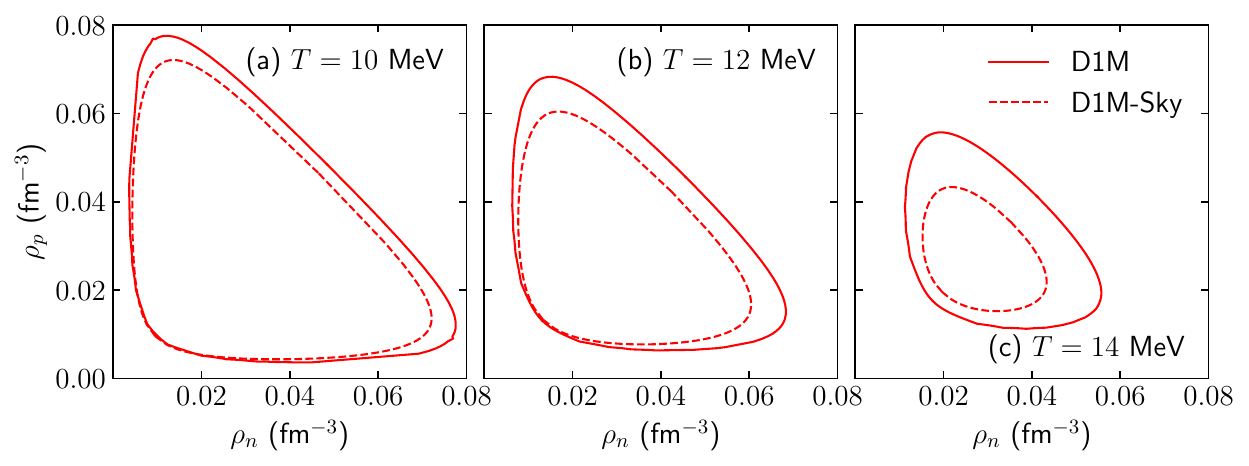}
\caption{{Spinodal contours at (a)~$T=10$~MeV, (b)~$T=12$~MeV, and (c)~$T=14$~MeV for D1M and D1M-Sky parametrizations.}}
\label{fig:D1Mcspi}
\end{figure*}

\begin{figure*}[!htb]
\centering
\includegraphics[scale=0.7]{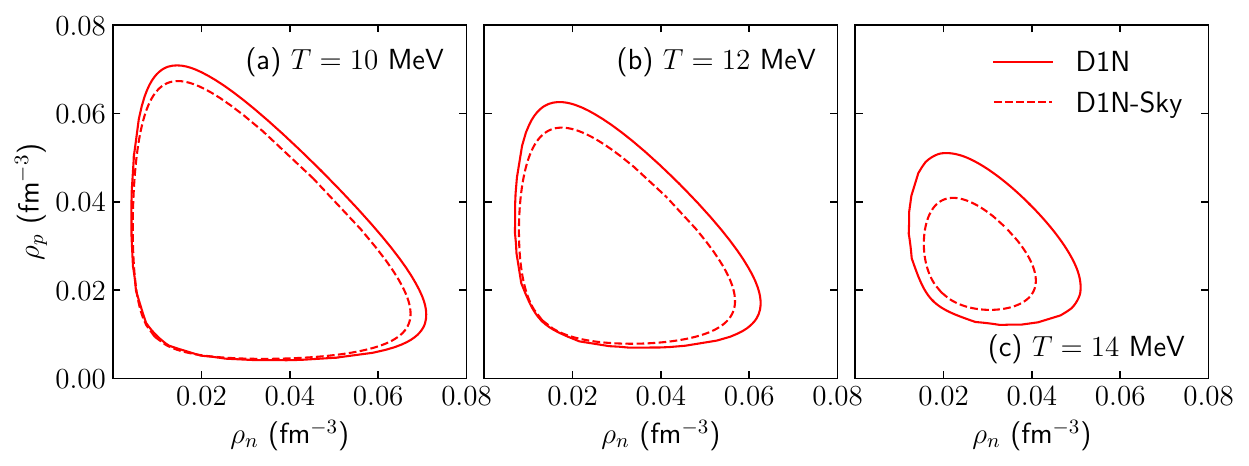}
\caption{{Spinodal contours at (a)~$T=10$~MeV, (b)~$T=12$~MeV, and (c)~$T=14$~MeV for D1N and D1N-Sky parametrizations.}}
\label{fig:D1Ncspi}
\end{figure*}

\begin{figure*}[!htb]
\centering
\includegraphics[scale=0.7]{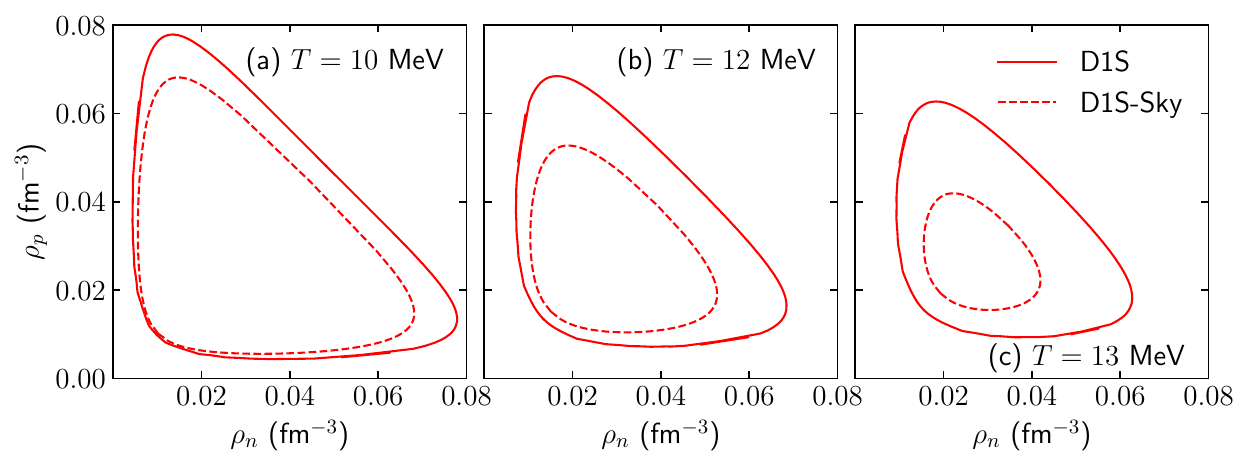}
\caption{{Spinodal contours at (a)~$T=10$~MeV, (b)~$T=12$~MeV, and (c)~$T=13$~MeV for D1S and D1S-Sky parametrizations.}}
\label{fig:D1Scspi}
\end{figure*}

At this point, we present the predictions of the Gogny parametrizations and their corresponding Skyrme-like versions regarding the unstable thermodynamical regions. We refer, more specifically, to the calculation of the chemical instability boundaries obtained through the condition
\begin{align}
\left(\frac{\partial\mu_q}{\partial y}\right)_{P,T} &= 0.
\label{eq:cspi}
\end{align}
Once again, we use the relation given in Eq.~\eqref{eq:dmudy} in order to convert the equation to be solved to the following one: $f(\rho,y,T) = 0$. In this case, for a fixed temperature we run the density and find the respective proton fraction and, consequently, the asymmetry parameter. By taking this procedure into account we are able to construct the spinodal contours in the neutron--proton density space for different temperatures, since by having $y$ for each $\rho$ we find $\rho_p=y\rho$ and $\rho_n=(1-y)\rho$. The results are plotted in Figs.~\ref{fig:D1Mcspi}-\ref{fig:D1Scspi}.
Notice in these figures the typical behavior exhibited by the instability (internal) region of each Gogny parametrization, namely, the shrinkage as the temperature increases. In the limiting case, each region collapses to a single point at the SNM critical temperature. By analyzing the zero-range versions of the Gogny model, it is clear that the respective spinodal contour is systematically smaller than the original one. As pointed out in Ref.~\cite{vidana2008} in a study performed with the Argonne V18 interaction~\cite{wiringa95} along with an Urbana-type three-body force in the context of the Brueckner--Hartree--Fock approximation, in comparison with Skyrme (SLy230a~\cite{chabanat97}) and relativistic mean
field models (NL3~\cite{lalazissis97} and TW~\cite{typel99}), in the SNM case the larger the critical temperature~($T_c^\sm$) and saturation density at zero temperature~($\rho_0$) are, the larger the instability (spinodal) region is. In the case of the Skyrme-versions constructed in this work, $\rho_0$ is the same as its corresponding Gogny model for each parametrization. However, as we have shown before, the results presented in Table~\ref{tab:snm} indicate that $T_c^\sm$ is always smaller for the Skyrme-like version than for the respective Gogny parametrization. Therefore, one can assign the smaller spinodal region in the Skyrme-like models to the smaller critical temperature.

\subsection{Isospin distillation}

\begin{figure*}[!htb]
\centering
\includegraphics[scale=0.6]{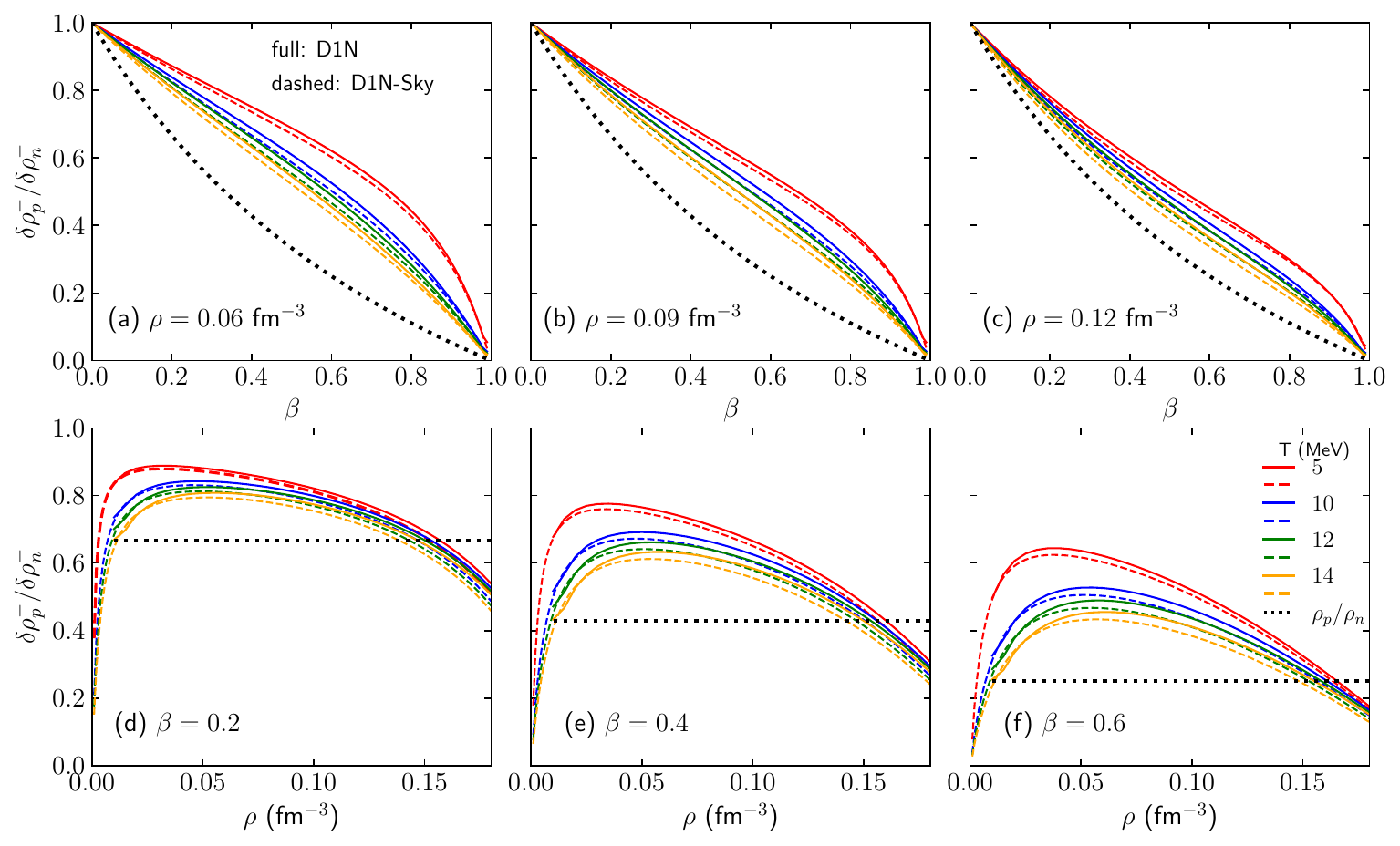}
\caption{{Ratio $\delta \rho_p^-/\delta \rho_n^-$ for the Gogny D1N model and its respective Skyrme zero-range version for several temperatures. Upper (lower) panels: results as a function of the asymmetry parameter (density) for fixed values of density (asymmetry parameter).}}
\label{fig:D1Ndest}
\end{figure*}

We also analyze here the stability of asymmetric matter against separation into two phases. The nuclear system, described by the Gogny model and its Skyrme pattern in our case, is considered stable if its single-phase free energy is lower than the one related to the two-phase configurations~\cite{margueron2003}. The consequence of such a condition is that the free energy density of the system is a convex function of $\rho_p$ and~$\rho_n$, resulting in a curvature matrix, given by
\begin{align}
[\mathcal{H}_{ij}] = \left[ \frac{\partial^2 \mathcal{H}}{\partial \rho_i \partial \rho_j}\Bigg|_T \right] = \left[ \frac{\partial \mu_j}{\partial \rho_i}\bigg|_T  \right],
\label{eq:mcurv}
\end{align}
being positive-definite and with the two associated eigenvalues,
\begin{align}
\lambda_{\pm} = \frac{1}{2} \left[ \text{Tr}(\mathcal{H}_{ij}) \pm \sqrt{\text{Tr}(\mathcal{H}_{ij})^2 - 4 \text{Det}(\mathcal{H}_{ij})} \right],
\end{align}
greater than zero as well. The respective eigenvectors are $(\delta\rho_p^{\pm},\delta\rho_n^{\pm})$, and the one associated with the lower eigenvalue $\lambda_-$ indicates the direction of the instability, specifically characterized in the case of nuclear matter by the ratio~\cite{margueron2003}
\begin{align}
\frac{\delta \rho_p^-}{\delta \rho_n^-} = \frac{\mathcal{H}_{np}}{\lambda_- - \mathcal{H}_{pp}} = \frac{\lambda_- - \mathcal{H}_{nn}}{\mathcal{H}_{np}}.
\label{eq:ratio}
\end{align}
In order to evaluate the matrix elements of Eq.~\eqref{eq:mcurv}, we proceed to rewrite the derivatives of $\mathcal{H}$ in terms of $\rho$ and~$y$, since $\mathcal{H}$ is given in terms of these quantities instead of $\rho_p$ and $\rho_n$.  The final expressions are
\begin{align}
\frac{\partial^2 \mathcal{H}}{\partial \rho_n^2} &=
\frac{y^2}{\rho^2} \frac{\partial^2 \mathcal{H}}{\partial y^2}
+ \frac{2y}{\rho^2} \frac{\partial \mathcal{H}}{\partial y}
- \frac{2y}{\rho} \frac{\partial^2 \mathcal{H}}{\partial y \partial \rho}
+ \frac{\partial^2 \mathcal{H}}{\partial \rho^2},
\end{align}
\begin{align}
\frac{\partial^2\mathcal{H}}{\partial \rho_p^2} &=
\frac{(1-y)^2}{\rho^2} \frac{\partial^2 \mathcal{H}}{\partial y^2}
- \frac{2(1-y)}{\rho^2} \frac{\partial \mathcal{H}}{\partial y}
\nonumber\\
&+ \frac{2(1-y)}{\rho} \frac{\partial^2 \mathcal{H}}{\partial y \partial \rho}
+ \frac{\partial^2 \mathcal{H}}{\partial \rho^2},
\end{align}
and
\begin{align}
\frac{\partial^2\mathcal{H}}{\partial\rho_p\partial\rho_n} &=
- \frac{(1 - y)y}{\rho^2}\frac{\partial^2\mathcal{H}}{\partial y^2}
+ \frac{(2y - 1)}{\rho^2}\frac{\partial\mathcal{H}}{\partial y}
\nonumber\\
&+ \frac{(1 - 2y)}{\rho}\frac{\partial^2\mathcal{H}}{\partial y\partial\rho}
+ \frac{\partial^2\mathcal{H}}{\partial\rho^2},
\end{align}
where we have considered $\partial^2\mathcal{H}/\partial y\partial\rho=\partial^2\mathcal{H}/\partial\rho\partial y$, assumption that leads to $\partial^2\mathcal{H}/\partial\rho_n\partial\rho_p=\partial^2\mathcal{H}/\partial\rho_p\partial\rho_n$. In Fig.~\ref{fig:D1Ndest} we present our results for the ratio given in Eq.~\eqref{eq:ratio} for some fixed values of $T$.
Since all Gogny parametrizations studied in this work give practically the same curves, we decided to exhibit only one of them, namely, the D1N model in particular.

As a general trend, we observe that the difference in $\delta \rho_p^-/\delta \rho_n^-$ between D1N and \mbox{D1N-Sky}, although not so large, increases with temperature, exactly as verified in the previous quantities analyzed in this work. Furthermore, one also notices that $\delta \rho_p^-/\delta \rho_n^-$ is always greater than $\rho_p/\rho_n$ in the case of fixed total density, see upper panels of Fig.~\ref{fig:D1Ndest}, indicating a more symmetric composition in the liquid phase of the asymmetric nuclear matter system in comparison with the gaseous phase, that is more neutron-rich, by particle conservation~\cite{vidana2008,dalen,margueron2003}. Such unbalance leads to an isospin distillation, or fragmentation effect, according to previous analysis in this regard, see for instance Refs.~\cite{vidana2008,dalen,margueron2003}. As a direct consequence, $\delta \rho_p^-/\delta \rho_n^-$ is considered as a ``measurement'' of the isospin symmetry restoration efficiency. In other words, greater (smaller) values of $\delta \rho_p^-/\delta \rho_n^-$ indicate greater (smaller) efficiency.

Notice that the Gogny model and its Skyrme-version present the same features concerning this effect, i.e., $\delta \rho_p^-/\delta \rho_n^-$ decreases with temperature and also with~$\beta$ (or, equivalently, increases with $y$). As a function of density for fixed asymmetry $\beta$ (see lower panels of Fig.~\ref{fig:D1Ndest}), on the other hand, it is clear that $\delta \rho_p^-/\delta \rho_n^-$ increases with density at a low-density regime, and then starts to decrease with $\rho$. These results are qualitatively very similar to other ones found in the literature. We address the reader to Ref.~\cite{vidana2008}, for instance, for a study of spinodal instabilities of asymmetric nuclear matter at finite temperature within the microscopic Brueckner--Hartree--Fock approach using the realistic Argonne V18 potential plus a three-body force of Urbana type. The ratio $\delta \rho_p^-/\delta \rho_n^-$ exhibits the same features as the ones presented here. In Ref.~\cite{dalen}, a density-dependent relativistic Hartree model based on a microscopic Dirac--Brueckner--Hartree--Fock approach, using the realistic Bonn-A potential, was used to also investigate the behavior of $\delta \rho_p^-/\delta \rho_n^-$. Our findings are also in agreement with the results coming from this particular analysis.

\section{Summary and concluding remarks}
\label{sec:summ}

We have analyzed in this work the predictions of the Gogny model, through the parametrizations of the D1 family in particular, in describing asymmetric nuclear matter at finite temperature. More specifically, we investigated the phase coexistence by means of the corresponding binodal sections and spinodal contours, and also the phenomenon of isospin distillation related to the instability regions. We also extended the calculation of the nucleon effective mass of the Gogny model to finite temperature and showed the temperature and density dependence of this quantity as a function of momentum in some selected examples. Furthermore, for each Gogny parametrizations (D1M, D1N, and D1S) we constructed a respective zero-range Skyrme version (\mbox{D1M-Sky}, \mbox{D1N-Sky}, and \mbox{D1S-Sky}) where the free constants of the Skyrme model were obtained in order to reproduce the same saturation density, binding energy, incompressibility, isoscalar effective mass, isovector effective mass, symmetry energy, and symmetry energy slope of the original Gogny model. These quantities were evaluated at $T=0$ and $\rho=\rho_0$. By taking this procedure into account, we estimate the impact of the finite range of the nuclear interaction in warm and asymmetric nonrelativistic nuclear matter. 

As a first analysis, we investigated the particular case of symmetric nuclear matter by calculating its critical parameters through the solution of the set of coupled equations related to the first and second derivatives of pressure with respect to density. As a result, we found that the values of $T_c$, $\rho_c$, and $P_c$ are systematically reduced for the Skyrme-like parametrizations, indicating that the heating of the system described by a zero-range model reaches a well-defined single thermodynamical phase before the system in which the Gogny model is applied. In addition to this study, we computed the binodal sections of the models in ANM. Our findings point out a shrinkage of the binodal sections obtained from the zero-range model in comparison with those evaluated through the Gogny model. Consequently, the Skyrme parametrizations collapse to a single point at $T=T_c$ of SNM faster than the respective Gogny partners. Due to this effect, both, the critical pressure of the Skyrme-like model and the pressure related to the point of maximum asymmetry are smaller than the respective quantities of the finite-range model. With regard to the unstable thermodynamical regions, analyzed by means of the chemical instability boundaries calculated from $(\partial\mu_q/\partial y)_{P,T}=0$, we also verified the trend that the spinodal contours of the zero-range model are located inside the ones of the original Gogny force.

Last but not least, we also analyzed the efficiency of the system concerning the isospin symmetry restoration. This quantity is directly connected to the ratio $\delta \rho_p^-/\delta \rho_n^-$, which, in turn, depends on the lower eigenvalue of the curvature matrix, whose elements are given by $[\mathcal{H}_{ij}] = [ (\partial^2 \mathcal{H})/(\partial \rho_i \partial \rho_j)|_T ]$, for $i,j=p,n$. For the sake of completeness, we provided such elements as derivatives of $\rho$ and $y$ since these are the natural quantities on which the Helmholtz free energy density depends. Our results show that $\delta \rho_p^-/\delta \rho_n^-$ is always greater than $\rho_p/\rho_n$ in the case of fixed total density. This feature indicates a more symmetric composition in the liquid phase than in the gaseous one (more neutron-rich) for both, Gogny and zero-range models. Actually, the asymmetry difference between the liquid and gas phases in coexistence can be seen already in Figs.~\ref{fig:D1Mbin}-\ref{fig:D1Sbin}. Furthermore, the difference between the Gogny and Skyrme-version models increases with temperature.

\section{Appendix}
In this appendix we provide details of the derivation of Eq.~\eqref{eq:mgogvec}. 

According to Refs.~\cite{pearson2001,baoanli2018}, the isovector effective mass is defined in such a way that the effective mass for each type of nucleon can be expressed as
\begin{align}
f_q = f_s + \tau_{3q} \beta (f_s - f_v),
\label{eq:fq}
\end{align}
where $f_q=M/M^*_q$ ($q=n,p$), $f_s=M/M^*_s$, and $f_v=M/M^*_{\mathrm{vec}}$. Here, $M^*_s$ and $M^*_{\mathrm{vec}}$ represent the isoscalar and the isovector effective masses, respectively, at arbitrary densities in symmetric nuclear matter.

In Skyrme models, $f_q$ is momentum-independent and is linear in the asymmetry parameter $\beta$, cf.\ Eq.~\eqref{eq:skyefmass},
and, consequently, the relation \eqref{eq:fq} is exact. 
For a Skyrme force it is straightforward to obtain the expressions for $f_s$ and $f_v$ 
from Eq.~\eqref{eq:skyefmass}, since $f_s$ is given by either $f_n$ or $f_p$ in SNM ($\beta=0$),
and $f_v$ is given by the expression of $f_p$ in pure neutron matter (as seen from Eq.~\eqref{eq:fq} at $\beta=1$).
One finds
\begin{align}
f_s=  1+\frac{M\rho}{4\hbar^2} (a + b) , \quad f_v = 1+\frac{M\rho}{4\hbar^2} a , 
\label{eq:skyfsfv}
\end{align}
in terms of the parameters $a$ and $b$ defined in Eq.~\eqref{eq:ab}.

In the case of finite-range models such as the Gogny models, however, the effective mass at $T=0$ is a more complicated
function of $k$, $k_{F_n}$ and $k_{F_p}$, i.e., $f_q = f_q(k,k_{Fq},k_{F\bar{q}})$, and  the
dependence of $f_q$ on $\beta$ is no longer linear. Still, Eq.~\eqref{eq:fq} can be applied to first order in $\beta$.
Normally, one focuses on the nucleon effective masses evaluated at $k= k_{Fq}$, i.e., at the Fermi momentum of the same nucleon
\cite{arnau1,davesne2025,baoanli2018,ring2023}. We assume this and expand $f_q$ to first order in $\beta$ for applying Eq.~\eqref{eq:fq}.
Taking into account that $k_{Fq}= k_F ( 1 + \tau_{3q} \beta)^{1/3}$, to linear order in $\beta$ it is
\begin{align}
k_{Fq} &= k_F \left( 1 + \tau_{3q} \frac{\beta}{3} \right)
\end{align}
and
\begin{align}
f_q = f_F + \frac{k_F \beta}{3} \left(\tau_{3q}\frac{\partial f_q}{\partial k_{Fq}} 
    + \tau_{3\bar{q}}\frac{\partial f_q}{\partial k_{F\bar{q}}} \right) \Bigg|_{k_{F}} ,
\end{align}
where we indicate that after performing the derivatives, they are calculated at $k_{Fq}= k_{F\bar{q}}= k_F$. Thus, one has
%
\begin{align}
f_q \pm f_{\bar{q}} &= f_F \pm f_F
+ \frac{k_F \beta}{3} \left(\tau_{3q}\frac{\partial f_q}{\partial k_{Fq}}
\pm       \tau_{3\bar{q}}\frac{\partial f_{\bar{q}}}{\partial k_{F\bar{q}}} \right) \Bigg|_{k_{F}} 
\nonumber \\
&+ \frac{k_F \beta}{3} \left(\tau_{3\bar{q}}\frac{\partial f_q}{\partial k_{F\bar{q}}}
\pm       \tau_{3q}\frac{\partial f_{\bar{q}}}{\partial k_{Fq}} \right) \Bigg|_{k_{F}} .
\end{align}
Taking into account that $\tau_{3\bar{q}} = -\tau_{3q}$ and the fact that $f_q$ of Gogny forces fulfils
\begin{align}
& \frac{\partial f_q}{\partial k_{Fq}} \Bigg|_{k_{F}}  = \frac{\partial f_{\bar{q}}}{\partial k_{F\bar{q}}} \Bigg|_{k_{F}} , \quad
 \frac{\partial f_q}{\partial k_{F\bar{q}}} \Bigg|_{k_{F}}  = \frac{\partial f_{\bar{q}}}{\partial k_{Fq}} \Bigg|_{k_{F}} ,
\end{align}
it is seen that
\begin{align}
f_q + f_{\bar{q}} &= 2 f_F ,
\nonumber \\
f_q - f_{\bar{q}} &=  2 \tau_{3q} \frac{k_F \beta}{3}  
\left(\frac{\partial f_q}{\partial k_{Fq}} - \frac{\partial f_q}{\partial k_{F\bar{q}}} \right) \Bigg|_{k_{F}} .
\label{eq:fqpm}
\end{align}
By using Eq.~\eqref{eq:fq} in \eqref{eq:fqpm}, one finds
\begin{align}
f_s = f_F
\end{align}
and
\begin{align}
f_s - f_v =
\frac{k_F}{3} \left(\frac{\partial f_q}{\partial k_{Fq}}
- \frac{\partial f_q}{\partial k_{F\bar{q}}} \right) \Bigg|_{k_{F} ,}
\end{align}
which allows, after a little algebra and by using Eq.~\eqref{eq:mgog} at $T=0$, to obtain for Gogny forces the result
\begin{align}
f_s &=
\nonumber\\
& 1 + \frac{2M}{\sqrt{\pi}\hbar^2}\sum_{i=1}^{2} \mu_i^2 \left[ M_i + \frac{H_i}{2} - \frac{B_i}{2} - \frac{W_i}{4} \right]F_3(\mu_ik_F) 
\label{eq:fs}
\end{align}
and the expression of the isovector effective mass given in Eq.~\eqref{eq:mgogvec}, namely,
\begin{align}
f_v &= 1 + \frac{2M}{\sqrt{\pi}\hbar^2}\sum_{i=1}^{2} \mu_i^2 \left[ 2M_i + H_i - B_i - \frac{W_i}{2} \right]F_3(\mu_ik_F)
\nonumber\\
&+ \frac{2M}{9\sqrt{\pi}\hbar^2}\sum_{i=1}^{2} \mu_i^2 \Bigg[ M_i + \frac{H_i}{2} - B_i - \frac{W_i}{2} \Bigg]F_4(\mu_ik_F),
\label{eq:fv}
\end{align}
where
\begin{align}
F_3(x) &= \frac{1}{x^3} \left[ 2 - x^2 - (2 + x^2)e^{-x^2} \right],
\\
F_4(x) &= 9\rho\frac{d^2 (\rho F_1(x))}{d \rho^2} = 3 \left[ \frac{1}{x} - \left( x + \frac{1}{x} \right)e^{-x^2} \right],
\end{align}
and
\begin{align}
F_1(x) &= \frac{\sqrt{\pi}}{2} \operatorname{erf}(x) + \left( \frac{1}{2x} - \frac{1}{x^3} \right) e^{-x^2} - \frac{3}{2x} + \frac{1}{x^3}.
\end{align}
Notice that in the expressions above one has
\begin{align}
\rho = \frac{2 k_F^3}{3 \pi^2}
\end{align}
and
\begin{align}
x = \mu_i k_F, \quad \frac{d}{d\rho} &= \mu_i \frac{d k_F}{d \rho} \frac{d}{dx} = \frac{x}{3 \rho} \frac{d}{dx} .
\end{align}

\begin{figure}[!htb]
\centering
\includegraphics[scale=0.32]{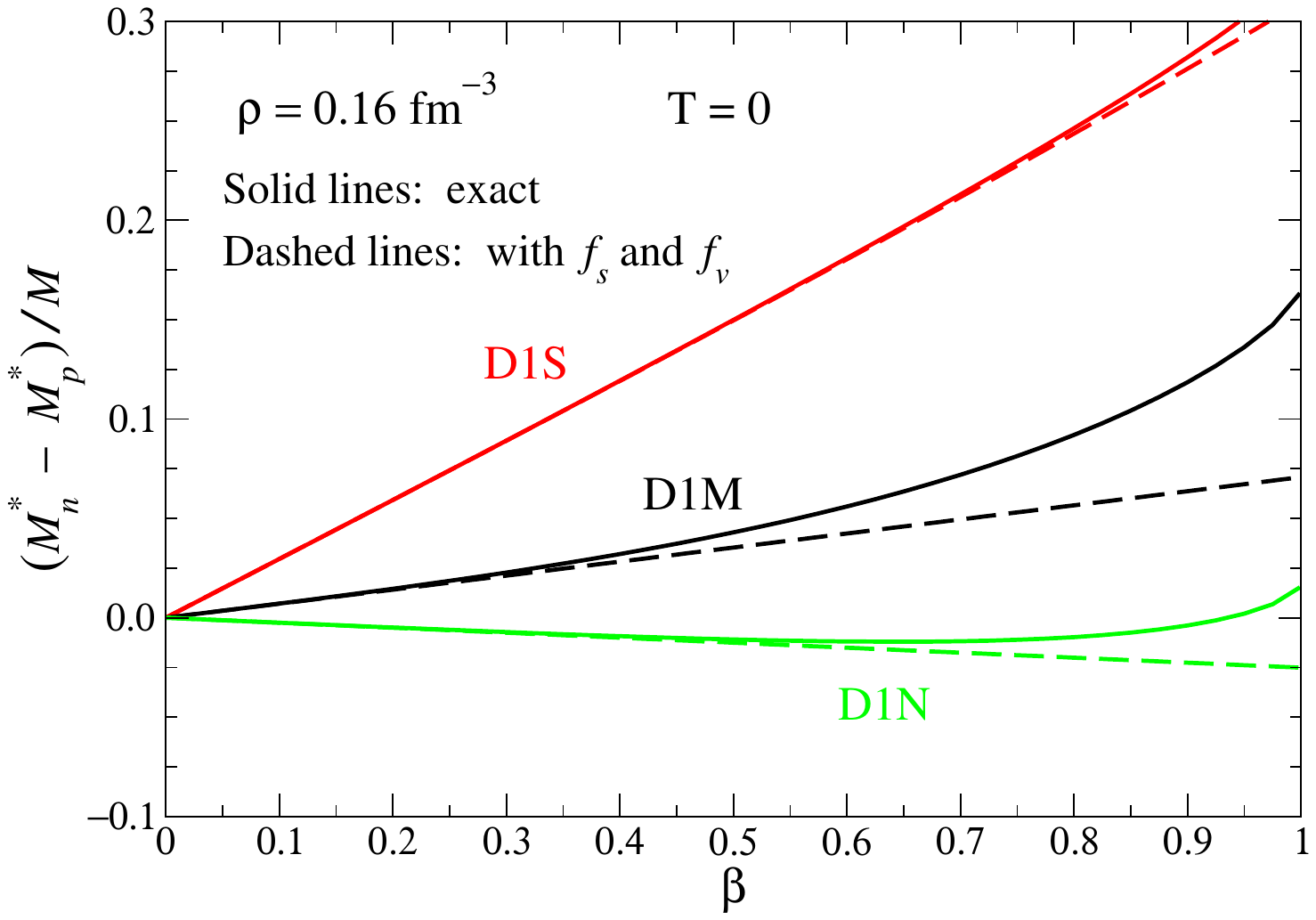}
\caption{Neutron-proton effective mass splitting of the Gogny models at $\rho=0.16$~fm$^{-3}$ and $T=0$ as a function of the asymmetry parameter $\beta$. The results of the dashed lines are calculated with Eq.~\eqref{eq:fq} using the isoscalar and isovector effective masses of the models.}
\label{fig:mstarn-mstarp}
\end{figure}

For illustration purposes, in Fig.~\ref{fig:mstarn-mstarp} we plot the neutron-proton effective mass splitting $(M^*_n-M^*_p)/M = f_n^{-1}(k=k_{F_n},k_{F_n},k_{F_p})-f_p^{-1}(k=k_{F_p},k_{F_p},k_{F_n})$ as a function of asymmetry $\beta$, at density $\rho=0.16$~fm$^{-3}$ and zero temperature, for the Gogny models considered in this paper. 
We show by dashed lines the same splitting calculated using Eq.~\eqref{eq:fq} for $f_n$ and $f_p$, with $f_s$ and $f_v$ given by Eqs.~\eqref{eq:fs} and \eqref{eq:fv}.

\begin{acknowledgements}
Useful discussions with A. Rios are appreciated. This work is a part of the project INCT-FNA proc. No. 464898/2014-5. It is also supported by Conselho Nacional de Desenvolvimento Cient\'ifico e Tecnol\'ogico (CNPq) under Grants No. 307255/2023-9 (O.L.), No. 308528/2021-2 (M.D.). O.~L. and M.~D. also thank CNPq for the project No. 01565/2023-8~(Universal). M.~C. and X.~V. acknowledge partial support from Grants No.\ PID2023-147112NB-C22 and No.\ CEX2019-000918-M (through the ``Unit of Excellence Mar\'{\i}a de Maeztu 2020-2023'' award to ICCUB) from the Spanish MCIN/AEI/10.13039/501100011033.
\end{acknowledgements}

\bibliographystyle{apsrev4-2}
\bibliography{references-revised}

\end{document}